\documentclass[usenatbib]{mnras}

\usepackage{graphicx}
\usepackage{times}

\usepackage{epstopdf}
\usepackage{float}
\usepackage{subcaption}

\newcommand{\sunrise}{{\footnotesize SUNRISE}}

\newcommand{\gadget}{{\footnotesize GADGET-3}}

\newcommand{\msun}{M$_{\odot}$}

\newcommand{\inv}{$^{-1}$}

\newcommand{\nh}{$N_{\rm H}$}

\newcommand{\jwst}{{\em JWST}}

\newcommand{\wise}{{\em WISE}}
\newcommand{\wonetwo}{{$ W1-W2$}}
\newcommand{\wtwothree}{{$ W2-W3$}}

\title[IR-selected AGN in mergers]{The power of infrared AGN selection in mergers: a theoretical study}

\author[L.~Blecha et al.]{
Laura Blecha, $^{1}$\thanks{E-mail: lblecha@ufl.edu}
Gregory F.~Snyder,$^2$
Shobita Satyapal,$^3$
and Sara L.~Ellison$^4$
\\
$^1$ University of Florida, Department of Physics, 2001 Museum Rd., Gainesville, FL 32611, USA\\
$^2$ Space Telescope Science Institute, 3700 San Martin Dr, Baltimore, MD 21218, USA\\
$^3$ George Mason University, Department of Physics \& Astronomy, MS 3F3, 4400 University Drive, Fairfax, VA 22030, USA\\
$^4$ Department of Physics \& Astronomy, University of Victoria, Finnerty Road, Victoria, British Columbia, V8P 1A1, Canada
}

\begin{document}
\maketitle

\begin{abstract}
The role of galaxy mergers in fueling active galactic nuclei (AGN) is still debated, owing partly to selection effects inherent to studies of the merger/AGN connection. In particular, luminous AGN are often obscured in late-stage mergers. Mid-infrared (IR) color selection of dust-enshrouded AGN with, e.g., the {\em Wide-field Infrared Survey Explorer} (\wise) has uncovered large new populations of obscured AGN. However, this method is sensitive mainly to AGN that dominate emission from the host. To understand how selection biases affect mid-IR studies of the merger/AGN connection, we simulate the evolution of AGN throughout galaxy mergers. Although mid-IR colors closely trace luminous, obscured AGN, we show that nearly half of merger-triggered AGN are missed with common mid-IR selection criteria, even in late-stage, gas-rich major mergers. At $z \la 0.5$, where merger signatures and dual nuclei can most easily be detected, we find that a more lenient \wonetwo\ $> 0.5$ cut greatly improves completeness without significantly decreasing reliability. Extreme nuclear starbursts are briefly able to mimic this AGN signature, but this is largely irrelevant in mergers, where such starbursts are accompanied by AGN. We propose a two-color cut that yields high completeness and reliability even in starbursting systems. Further, we show that mid-IR color selection very effectively identifies {\em dual} AGN hosts, with the highest fraction at the smallest separations ($< 3$ kpc). Thus, many merger hosts of mid-IR AGN should contain unresolved dual AGN; these are ideal targets for high-resolution follow-up, particularly with the {\em James Webb Space Telescope}.
\end{abstract}


\section{Introduction}
\label{sec:intro}

Mergers between galaxies cause morphological transformations, induce bursts of star formation, and fuel active galactic nuclei (AGN), and the resulting stellar and AGN feedback regulates further evolution. Galaxy mergers therefore offer a plausible mechanism for coordinated evolution between supermassive black hole (BHs) and their host galaxies, giving rise to the observed correlations between BH mass and stellar bulge properties \citep[e.g.,][]{magorr98,gebhar00,gultek09,mccma13}. However, despite a great deal of study, the connection between galaxy mergers and AGN fueling is still not well understood.

The large majority of AGN hosts do not show signs of ongoing merger activity, suggesting that secular mechanisms are responsible for triggering most AGN and leading some to conclude that there is no convincing evidence for a merger/AGN connection \citep[e.g.,][]{cister11, kocevs12, schawi12, villfo14, villfo17}. In contrast, studies that select samples of paired and merging galaxies have found that when mergers do occur, they are more likely than isolated galaxies to host AGN, and that AGN are most likely to be found in the late stages of the merger \citep[e.g.,][]{liu11a, elliso11, silver11, elliso13b, satyap14, lackne14, weston17, gouldi18}.  The merger/AGN association also depends strongly on the AGN luminosity \citep[e.g.,][]{elliso13b,satyap14,koss12}. At the highest quasar luminosities, {\em most} AGN may be hosted in ongoing mergers \citep[e.g.,][]{urruti08,fan16,vito18}. 

Crucially, the observed correlation between mergers and AGN fueling is highly sensitive to obscuration by gas and dust, which can be extreme during merger-induced nuclear starbursts. Indeed, the merger fraction of AGN hosts is much higher for infrared (IR) and hard X-ray selected AGN, which are less sensitive to dust obscuration, than for optical or soft X-ray selected AGN \citep[e.g.,][]{sander88a, sander88b, sanmir96, veille09, koss10, satyap14, kocevs15}. Low-excitation radio galaxies, in contrast, have no apparent connection to galaxy mergers \citep{elliso15}. Ultraluminous infrared galaxies (ULIRGs, $L_{\rm IR} > 10^{12}$ L$_{\odot}$) emit most of their bolometric luminosity in the IR; in the nearby Universe, nearly all of these are strongly merging systems that contain buried starbursts or AGN \citep[e.g.,][]{sander88a, veille02, taccon08}. The most AGN-dominated ULIRGs are found in the latest stages of merging, with the highest IR luminosities \citep{veille09}.

The presence of dust in the nuclear region is a nearly ubiquitous feature of AGN. Partial obscuration by a dusty ``torus" (or another geometry with covering fraction less than unity) is commonly invoked in AGN unification theories to explain the Type I/Type II dichotomy observed in optical AGN spectra \citep[e.g.,][]{urrpad95}. Dusty AGN produce distinctive spectral signatures in the mid-IR. The hard ionizing AGN continuum emission heats the surrounding dust, up to the sublimation limit ($\sim$ 1500 K); the reprocessed emission appears as a red power-law slope in the mid-IR spectral energy distribution (SED; defined as $\sim 3$ - $30 \mu$m). This fact has been exploited to identify AGN based on their characteristic red mid-IR colors \citep[e.g.,][]{lauren00, lacy04, stern05, assef10, donley12}. Large new populations of AGN, most undetected in optical surveys, have been identified via mid-IR color selection in wide-field surveys, most recently the {\em Wide-field Infrared Survey Explorer} (\wise) all-sky survey \citep[e.g.,][]{jarret11, stern12, mateos12, mateos13, assef13, secres15}. 

The \wise\ bands ($W1$, $W2$, $W3$, and $W4$), are centered at 3.4, 4.6, 12, and 22 $\mu$m, respectively. \citet{stern12} proposed a single-color cut of \wonetwo\ $>0.8$ (Vega magnitudes) to identify luminous AGN with high reliability and completeness. Others have proposed two-color cuts to minimize contamination from star-forming galaxies, especially at $z \ga 1$ (\citealt[][hereafter J11]{jarret11}; \citealt{mateos12}; \citealt{assef13}). \citet{mateos12} also demonstrate that the completeness is a strong function of AGN luminosity, with high completeness achieved only for the most luminous AGN ($L_{\rm 2-10keV} > 10^{44}$ erg s\inv).

\citet{satyap14} used \wise\ color selection to identify a new population of low-redshift merger-triggered AGN. Using samples of merging and isolated selected from SDSS  \citep[][]{elliso11,elliso13b}, they cross-matched with the \wise\ data and found a significant excess of AGN activity in advanced mergers, relative to isolated galaxies and to the excess measured for optically-selected AGN.  The mid-IR selected AGN excess in mergers persists whether a selection cut of \wonetwo\ $>0.8$ or a more lenient \wonetwo\ $>0.5$ cut is used; this sample is at low redshift ($z<0.2$) and is therefore less sensitive to contamination from star-forming galaxies. Recently, \citet{weston17} and \citet{gouldi18} have confirmed the excess of \wise\ AGN in merging galaxies, and \citet{donley18} have found similar results for {\em Spitzer}-IRAC selected AGN (especially compared with 2-10 keV X-ray selected AGN) in the CANDELS/COSMOS field.

Similar trends are seen for hard X-ray selected AGN in merging galaxies \citep[e.g.,][]{koss10,koss12,kocevs15,ricci17,lansbu17,koss18}. Hard X-ray AGN selection is highly accurate and uniquely complete, but only to the comparatively shallow depth of current surveys with, e.g., {\em Swift}-BAT and {\em NuSTAR}. Mid-IR color selection yields much larger AGN samples, but it is sensitive only to AGN that are luminous relative to their host galaxies. Thus, to fully utilize mid-IR selection as a probe of the merger/AGN connection, we must understand how the selection completeness varies throughout the course of a merger and in different merger environments.

With this motivation, we use hydrodynamics and radiative transfer simulations of merging galaxies to model the evolution of nuclear obscuration and mid-IR SEDs through different merger stages. A similar approach to studying mid-IR AGN signatures in mergers has been used by \citet{snyder13} and \citet{roebuc16}, who focus on signatures sensitive to the AGN versus starburst contribution in ULIRGs. \citet{snyder13} find that AGN signatures can be suppressed by dust self-absorption in extreme starbursting $z\sim2$ galaxies, and they propose a combination of spectral signatures accessible to \jwst\ that can constrain the AGN fraction. \citet{roebuc16} compare the empirical and simulated AGN fraction in ULIRGs and find that the AGN fraction may be underestimated by empirical classifications in some cases.

In this work, we focus on the selection of merger-triggered AGN with \wise; we wish to quantify the completeness and accuracy of mid-IR AGN selection and how this depends on the properties of the merging system. Our other primary goal is determining the incidence of (possibly unresolved) close {\em dual} AGN in mid-IR selected samples -- i.e., simultaneously active BH pairs with $\la 1$ - 10 kpc separations. A small but growing sample of dual AGN has been identified in optically-selected surveys \citep[e.g.,][]{comerf09a, liu10a, comerf12, muller15}, but hard X-ray selected samples have revealed a much higher incidence of dual AGN in late-stage mergers \citep{koss12}. Follow-up studies of \wise\ AGN in mergers have also seen a high success rate in finding candidate dual AGN \citep{satyap17,elliso17}. We aim to determine the intrinsic dual AGN fraction in \wise-selected sources from our models, as a function of nuclear separation and AGN luminosity. Because we are interested specifically in mergers, we focus on modeling galaxies representative of relatively low-redshift systems ($z\la0.5$), similar to the observed samples of \citet{satyap14},\citet{weston17}, \& \citet{gouldi18}. At higher redshifts ($z \sim 1$-2) more typical of the overall \wise\ AGN population, merger signatures and dual AGN are increasingly difficult to reliably detect.

We use Vega magnitudes throughout this paper. The hydrodynamics simulations and radiative transfer calculations are described in \S~\ref{ssec:merger_sims} \& \ref{ssec:rt}, respectively.
Our results are presented in \S \ref{sec:results}.
Merger-driven AGN obscuration is discussed in \S~\ref{ssec:obscuration}, and AGN lifetimes as a function of merger phase are presented in \S\ \ref{ssec:lifetimes}. We explore the completeness and reliability of mid-IR AGN selection in \S~\ref{ssec:completeness} and \ref{ssec:reliability}. In \S~\ref{ssec:dual_agn}, we consider the effectiveness of mid-IR selection of dual AGN. 
Finally, in \S~\ref{sec:conclusions} we discuss the implications of our results and summarize our main conclusions. 

\section{Methodology}
\label{sec:methods}

\subsection{Hydrodynamic Galaxy Merger Simulations}
\label{ssec:merger_sims}

As the basis for modeling AGN in mergers, we conduct high-resolution simulations of merging galaxies with \gadget, a smoothed-particle hydrodynamics (SPH) and N-body code that conserves energy and entropy and uses sub-resolution physical models for radiative heating and cooling, star formation, supernova feedback, metal enrichment, and a multi-phase interstellar medium \citep[ISM;][]{sprher03, spring05a}. BHs are modeled as gravitational ``sink" particles that accrete gas via an Eddington-limited, Bondi-Hoyle like prescription. Thermal AGN feedback is included by coupling 5\% of the accretion luminosity ($L_{\rm bol} = \epsilon_{\rm rad} \dot M c^2$) to the surrounding gas  as thermal energy, with a accretion-dependent radiative efficiency $\epsilon_{\rm rad}$ at low accretion rates \citep[cf.][]{narmcc08}. \gadget\ has been used for a multitude of studies of merging galaxies, including many studies of BH/galaxy co-evolution \citep[e.g.,][]{dimatt05, robert06b, cox06, hopkin06a, hopkin08c, blecha11, blecha13b}.

\begin{table}
\begin{center}
\begin{tabular}{c c c c c}
Name & $M_{\rm tot}$ & $M_*$ & $f_{\rm gas}$ & B/T \\ 
& [$10^{11}$ \msun] &[$10^{10}$ \msun] & & \\ \hline
A0  & 14 & 3.9 & 0.3 & 0 \\ 
B0  & 14 & 5.0 & 0.1 & 0 \\
C0 & 14 & 4.1 & 0.3 & 0.1 \\
D0& 14 & 4.2 & 0.3 & 0.2 \\
E0  & 14 & 5.1 & 0.1 & 0.2 \\ 
A1 & 6.8 & 2.0 & 0.3 & 0\\
B1 & 6.8 & 2.5 & 0.1 & 0 \\ 
C1  & 6.8 & 2.0 & 0.3 & 0.1 \\
D1 & 6.8 & 2.1 & 0.3 & 0.2 \\
E1  & 6.8 & 2.6 & 0.1 & 0.2 \\ 
A2 & 2.7 & 0.78 & 0.3 & 0 \\ \hline
\end{tabular}
\end{center}
\caption{Progenitor galaxy parameters. Column 1: Name of the progenitor used throughout the text. Columns 2 \& 3: Total and stellar mass of the galaxy, respectively. Column 4: Initial gas fraction in the galaxy disk (by mass, $M_{\rm gas}/M_{\rm *,disk}$). Column 5: Bulge-to-total mass ratio.  \label{table:galaxy_models}}
\end{table}

\begin{table}
\begin{center}
\setlength\tabcolsep{4 pt}
\begin{tabular}{c c c c c c c}
\# & Name & $q$ & SFR$_{\rm max}$ & sSFR$_{\rm max}$ & $N_{\rm H,max}$ & $L_{\rm AGN,max}$ \\
 & & & [\msun\ yr\inv] &[log yr\inv] & [log cm$^{-2}$] & [log erg s\inv] \\ \hline
0& {\bf A0A0} & 1 & 380 & -8.4 & 24.7 & 46.0 \\ 
1& {\bf A1A0} & 0.5 & 220 & -8.5 & 24.5 & 46.3\\
2& C1C0  & 0.5 & 14 & -9.7 & 23.6 & 45.2\\
3& {\bf D1D0} & 0.5 & 25 & -9.5 & 24.2 & 45.6 \\
4& {\bf A1E0} & 0.5 & 46 & -9.2 & 24.4 & 45.7 \\
5& A2A0  & 0.2  & 15 & -9.6 & 23.8 & 45.0 \\
6& B1B0  & 0.5  & 21 & -9.6 & 24.0 & 45.0 \\ 
7& E1E0  & 0.5  & 1.6 & -11 & 22.4 & 43.7 \\ \hline
\end{tabular}
\end{center}
\caption{Merger simulation parameters and key characteristics. Column 1: Simulation number. Column 2: Simulation name, indicating the combination of merging progenitor galaxies as in the notation of Table \ref{table:galaxy_models}.  Simulations names in boldface have also been run at ten times higher mass resolution. Column 3: Merging galaxy mass ratio. Columns 4 \& 5: Maximum total SFR and specific SFR (SFR/$M_*$) achieved during the simulation. Column 6: Maximum column density (along sight lines to either BH) achieved during the simulation, calculated as the median over all viewing angles in each snapshot. Column 7: Maximum bolometric AGN luminosity achieved during the simulation. \label{table:merger_models}}
\end{table}

As mentioned above, our primary goal is to study mid-IR AGN selection in merging galaxies, as probes of dual AGN and the merger/AGN connection. Identifying merging galaxies or dual AGN empirically requires the ability to resolve dual nuclei and to detect low-surface-brightness signatures of morphological disturbance (e.g. tidal tails). Because such signatures are increasingly difficult to identify at higher redshifts, our simulations are designed to be representative of galaxy mergers at relatively low redshift, rather than very gas rich ($f_{\rm gas} \ga 0.5$), extreme starbursting systems that may be more typical of some merger-triggered \wise\ AGN at the median redshift of the \wise\ AGN population ($z \sim 1$). This choice is in agreement with several recent empirical studies of IR AGN signatures in merging galaxies, which have similarly focused on the low redshift regime \citep{satyap14, weston17, gouldi18}.

Our primary simulation suite includes eight merger simulations with galaxy mass ratios of $q =$ 0.2 - 1; most of these are major mergers with $q=0.5$, but we include one equal-mass merger and one ``minor" merger with $q=0.2$. The galaxies consist of a dark matter halo, a disk of gas and stars, a stellar bulge, and a central BH.  We focus on merger progenitors with disk-dominated morphologies initially, choosing the bulge-to-total stellar mass ratio, B/T, between 0 (i.e., a pure disk initially) and 0.2. The initial gas fraction in the disk (by mass, $M_{\rm gas}/M_{\rm *,disk}$) is set to be 0.3 or 0.1, to represent both gas-rich and relatively gas-poor mergers. The fiducial baryonic gravitational softening length and mass resolution are $\epsilon_{\rm grav} =48$ pc and $m_{\rm b} = 2.8\times 10^5$ \msun, respectively. We also run a subset of these simulations with ten times higher mass resolution and $10^{1/3}$ times smaller softening length, $\epsilon_{\rm grav} = 23$ pc. A comparison between the fiducial- and high-resolution results is presented in Appendix \ref{ssec:resolution}.

The parameters of the progenitor galaxy models are summarized in Table \ref{table:galaxy_models}. The merger simulation parameters are given in Table \ref{table:merger_models}, along with some key characteristics of each merger, and the simulations that have also been run at higher resolution are denoted in boldface. Throughout the paper, we refer to the mergers either by a name, composed of the names assigned to each progenitor galaxy (e.g., A1A0), or by the simulation number (0-7) as listed in Table \ref{table:merger_models}. We also frequently refer to simulations 0-5 as the subset of ``gas-rich" mergers, simulations 0-4 as the ``gas-rich, major" mergers, and simulations 6-7 as the ``gas-poor" mergers.

\subsection{Dust Radiative Transfer Simulations}
\label{ssec:rt}

We conduct radiative transfer simulations in post-processing with the 3-D, polychromatic, Monte Carlo dust radiative transfer code {\footnotesize SUNRISE} \citep{jonsso06,jonsso10}. This publicly-available code has been used extensively with \gadget\ to model a wide range of isolated and merging galaxy populations  \citep[e.g.,][]{lotz11, narayanan10, haywar11, haywar13, snyder13, blecha13a, lanz14}. Stellar emission is calculated from single stellar population {\footnotesize STARBURST99} SEDs \citep{leithe99} based on the age and metallicity of each stellar particle, assuming a Kroupa initial mass function \citep{kroupa02}. Emission from HII regions and photodissociation regions (PDRs) around young stars is calculated using the {\footnotesize MAPPINGSIII} models of \citet{groves08}, where age, metallicity, and gas pressure are taken from the newly formed stellar particles in the \gadget\ simulation. A covering fraction of 0.2 is assumed for the PDR models, which include dust re-emission as well as polycyclic aromatic hydrocarbon (PAH) absorption and emission. 

An AGN SED is implemented based on the BH accretion rate using the luminosity-dependent templates of \citet{hopkin07a}. In the mid-IR, the AGN SED is derived from the empirically-determined mean quasar SEDs of \citet{richar06}, such that it implicitly includes reprocessed radiation from a sub-resolution dusty ``torus" typical of Type I quasars.

\begin{figure*}
\begin{subfigure}{0.45\textwidth}
\centering
\includegraphics[width=0.45\textwidth]{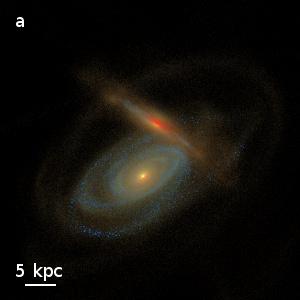}
\includegraphics[width=0.45\textwidth]{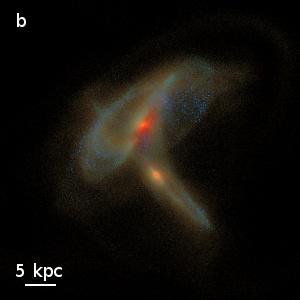}
\includegraphics[width=0.45\textwidth]{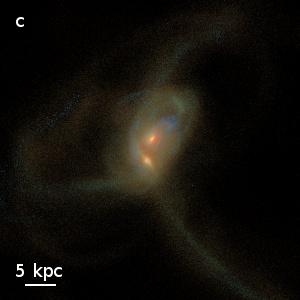}
\includegraphics[width=0.45\textwidth]{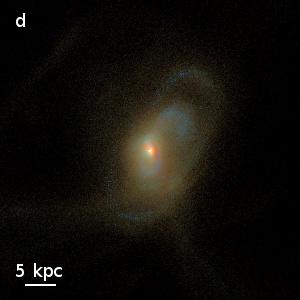}
\end{subfigure} 
\begin{subfigure}{0.45\textwidth}
\includegraphics[width=0.9\textwidth]{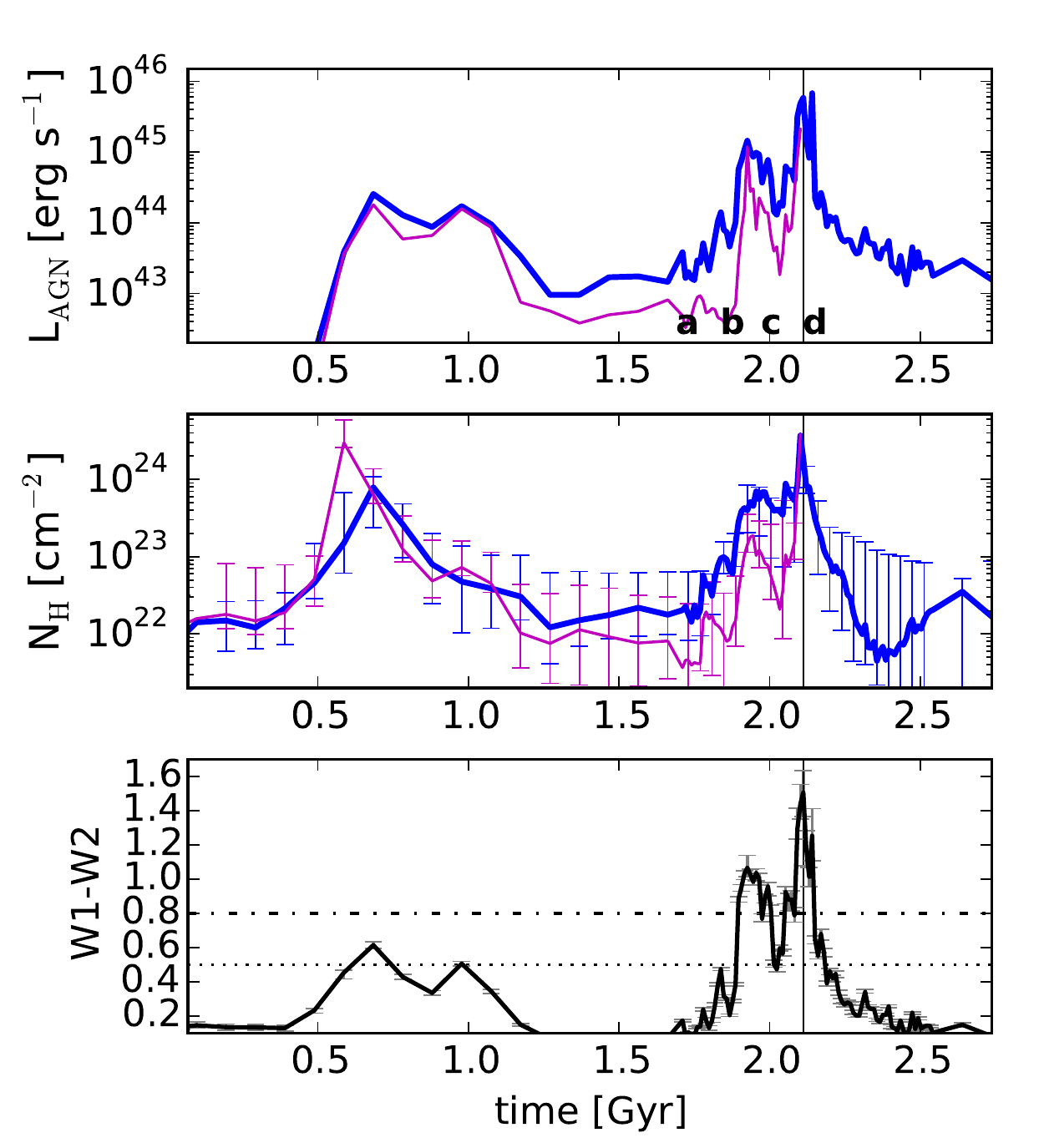}
\end{subfigure}
\caption{Evolution of an obscured AGN in the high-resolution simulation of the A1A0 merger. {\em Left panels:} SDSS $ugz$ images, generated from {SUNRISE} simulations, show the disturbed merger morphology during the late stages of the merger. In the first snapshot, the BH separation has just fallen below 10 kpc, and the final snapshot occurs just after the BH merger. The $ugz$ filter combination is chosen to enhance contrast between the dust-obscured nuclei and star-forming regions. {\em Right panels:} from top to bottom, the evolution of the bolometric AGN luminosity $L_{\rm AGN}$, line-of-sight gas column density $N_{\rm H}$, and \wise\ \wonetwo\ color throughout the merger is shown. The vertical line denotes the time of BH merger. In the $L_{\rm bol}$ and $N_{\rm H}$ plots, the blue and magenta curves correspond to each BH prior to merger, and the blue curve shows the post-merger evolution. In the $N_{\rm H}$ and \wonetwo\ plots, the error bars show the range of values over all viewing angles. $N_{\rm H}$ is calculated along the line of sight to each BH, in an aperture 64 pc in size (consistent with the effective spatial resolution of the high-resolution simulations). \wise\ colors are calculated for the entire galaxy. For clarity, error bars are plotted for only a subset of snapshots. The dashed and dotted lines in the \wonetwo\ plot denote single-color cuts of 0.5 and 0.8, used in the literature and in this work. Note that higher time resolution is used for the {SUNRISE} calculations in the late phases of the merger. The AGN luminosity and column density peak during the galaxies'  coalescence, when the galaxies are morphologically disturbed; this luminous, obscured AGN phase is closely traced by red \wonetwo\ colors. \label{fig:evol_with_images}} 
\end{figure*}

\begin{figure}
\begin{center}
\includegraphics[width=0.45\textwidth,  trim={0.1 0cm 0.1cm 0.75cm},clip]{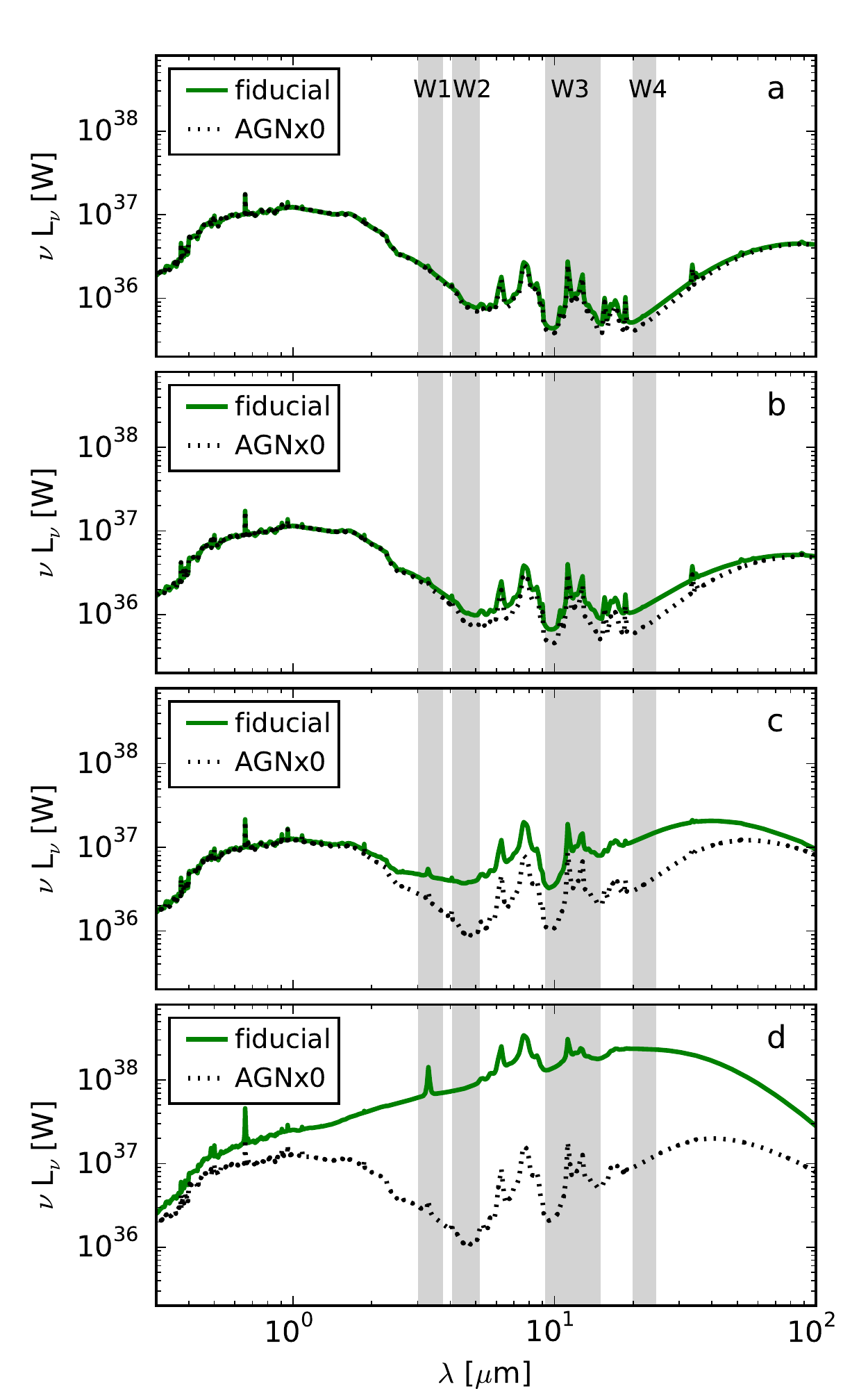}
\caption{Optical and IR SEDs are shown for the high-resolution A1A0 simulation snapshots corresponding to the optical images in Figure \ref{fig:evol_with_images}. Green solid lines give the fiducial simulation SED, and black dotted lines give corresponding SED in the AGNx0 simulation. The gray shaded regions denote the four \wise\ bands (from left to right: $W1$, $W2$, $W3$, \& $W4$). In the first two snapshots, the system is still dominated by star formation, and prominent PAH and 9.7 $\mu$m silicate absorption features are apparent. In snapshots ``c" and ``d" we see the reddening of the mid-IR colors with increasing AGN contribution, particularly in the last snapshot. The \wonetwo\ colors in each snapshot are 0.1 (``a"), 0.2 (``b"), 0.8 (``c"), and 1.3 (``d").  \label{fig:seds}}
\end{center}
\end{figure}

The dust distribution is calculated in \sunrise\ by projecting the gas-phase metal density onto a 3-D adaptively-refined grid, where we assume that 40\% of the metals are in dust \citep{dwek98} and use the Milky Way $R_{\rm V} = 3.1$ dust model of \citet{drali07}. We also tested the effect of using the \citet{drali07} dust model based on the Small Magellanic Cloud bar, which has fewer carbonaceous grains and hence weaker PAH features, but found that this did not significantly change our results. 

\gadget\ uses a multi-phase model for the ISM, in which cold gas is assumed to be clumpy, with a negligible volume filling factor. We make the same assumption about the gas (and hence the dust) in our \sunrise\ models: attenuation and emission from dust in cold gas clouds is neglected in the radiative transfer calculations, including the IR emission. \citep[This is equivalent to the ``multiphase-on" model of][]{haywar11}. Because galactic-scale dust primarily emits at far-IR wavelengths, ignoring this cold-phase dust emission should not significantly affect the mid-IR diagnostics considered here. In high-redshift, extremely gas-rich mergers where molecular gas dominates the ISM, the zero-volume-filling assumption for cold phase gas would break down, and a more appropriate choice would be to assume the gas and dust is distributed uniformly across the grid cells \citep[cf.][]{haywar11,snyder13}. The galaxies considered in this work are instead designed to be low-redshift analogues with a maximum initial gas fraction of 0.3, which are dominated by hot-phase gas throughout the merger. The most gas-rich of these do briefly become ultraluminous infrared galaxies (ULIRGs) with $L_{\rm IR} \sim 10^{12}$ L$_{\odot}$ during the peak of merging activity, with starburst and AGN activity driven by an increase in the cold gas fraction in the galactic nuclei. 

Using similar \sunrise\ simulations, \citet{snyder13} demonstrate that even the mid-IR signatures of AGN can be obscured by dust self-absorption in the nuclei of extreme starbursting, merging systems, but they find significant suppression of AGN signatures only for merger events featuring highly gas-rich $z \sim 2$ analogues that reach ``hyper-LIRG" luminosities ($L_{\rm IR} \sim 10^{13}$ L$_{\odot}$). For the less extreme, less obscured systems considered here, the multiphase ISM treatment that assumes negligible absorption and emission by dust in cold clumps is considered to be the most appropriate choice. Nonetheless, it is important to bear in mind that even mid-IR selection can miss the most heavily buried, Compton-thick AGN during peak obscuration in mergers, and that even in less-extreme mergers, select sight lines that intersect with cold gas clumps could obscure the AGN signatures in the mid-IR.

One should also keep in mind that these dust and ISM models are approximations of physics on sub-grid scales, an unavoidable consequence of the finite resolution of simulations. Dust grain distributions and gas clumpiness in the ISM of real galaxies will be spatially-dependent, evolving, and correlated with processes such as starbursts and AGN heating. The inherent uncertainty in sub-resolution dust and ISM models is an important caveat to any such study. The model parameters considered here are physically motivated and empirically based, and as noted above, we focus on a regime in which the primary concern of this study -- mid-IR color selection of AGN -- should not be subject to dramatic effects from these model uncertainties. 

Once the source SEDs are determined for stellar particles, star-forming regions, and AGN, and the dust distribution is determined from the gas grid, \sunrise\ performs Monte Carlo radiative transfer through the grid, computing energy absorption (including dust self-absorption) and thermal re-emission to produce spatially-resolved UV-to-IR SEDs. To ensure convergence of the dust temperature and IR emission calculation, the emergent IR luminosity in each grid cell is required to be converged to within a factor of 0.2 (such that the {\em integrated} IR SED is converged to a much higher degree). We find that adjusting this tolerance parameter between 0.1 and 0.4 has a negligible effect ($< 3\%$) on the emergent IR SED.

For each merger simulation, we run \sunrise\ on snapshots at 100 Myr intervals in the early stage of the merger and at 10 Myr intervals in the later stage of the merger. Images and resolved spectra are produced for seven isotropically-distributed viewing angles, with a spatial resolution of 500 pc or 167 pc in the early or late merger stages, respectively. (A $3\times$ larger field of view is used in the early merger phase, when the galaxies are at larger projected separations.) Unless otherwise specified, all results are presented for rest-frame SEDs, but we have also calculated broadband mid-IR magnitudes for SEDs redshifted up to $z=1$ for the fiducial simulation suite, and up to $z=4$ for select simulation snapshots. As discussed below, we find negligible difference in our results for $z \la 0.5$. 
 
Finally, in order to quantify the contribution of the AGN to the emergent SED, we also re-run each \sunrise\ simulation with the AGN luminosity artificially set to zero. We refer to these as the ``AGNx0" simulations. While this doesn't remove the effect of AGN feedback on the surrounding gas (because the \gadget\ simulation is the same for the fiducial and AGNx0 \sunrise\ simulations), it allows us to explicitly separate the contribution of stellar and AGN emission to the emergent SED. Among other things, this is essential for measuring the contribution of star-forming regions to the mid-IR colors.

\section{Results}
\label{sec:results}

The evolution of a high-resolution, gas-rich, major merger simulation is shown in Figure \ref{fig:evol_with_images}.\footnote{We use data from one of the high-resolution simulations here, to illustrate the morphological merger signatures in greater detail, but the qualitative and quantitative results are very similar for the fiducial-resolution simulations used for most of our analysis. See Appendix \ref{ssec:resolution} for a  comparison of key results between fiducial and high-resolution simulations.} The bolometric AGN luminosity and nuclear column density ($N_{\rm H}$, calculated along the line of sight to each BH\footnote{This aperture for the $N_{\rm H}$ calculation is chosen to be consistent with the size scale ($2.8\times \epsilon_{\rm grav}$) below which gravitational forces are mediated by the softening kernel. This quantity, which is 64 pc for the high-resolution simulations (as in Figure \ref{fig:evol_with_images}) and 136 pc for the fiducial simulations, can be considered the effective spatial resolution of the simulations.}) peak during final coalescence of the galaxies, supporting the idea that luminous, obscured AGN are preferentially triggered in late-stage galaxy mergers. The \wise\ mid-IR colors (which are calculated globally for the entire galaxy) closely trace this luminous, obscured AGN phase, with \wonetwo\ $> 0.8$ for 180 Myr during the late stages of the merger (defined from the time when the BH separation falls below 10 kpc, to 100 Myr after the BH merger). This corresponds to 46\% of the total late-stage merger phase. 

By the time of BH merger, most of the galaxies' initial gas content has been consumed by star formation (and BH accretion). Shortly after the BH merger, AGN feedback efficiently heats up and removes much of the remaining cold gas reservoir from the nucleus of the merger remnant, causing a precipitous decline in AGN luminosity, central gas column density, and star formation rate. Such AGN feedback episodes provide a means of regulating growth of both the galaxy bulge and the central BH \citep[e.g.,][]{wyiloe03,dimatt05,hopkin06a}, and they input substantial energy and metal-enriched gas into the circum-galactic medium \citep[e.g.,][]{hani18}. In the corresponding simulated optical images (Figure \ref{fig:evol_with_images}), significant reddening of the nuclear regions is apparent. Disturbed morphological merger signatures are also seen throughout the late merger phase. It is worth noting, however, that the most prominent tidal features have already begun to fade in the last image, just 30 Myr after the BH merger, when the AGN is at its peak luminosity. 

Figure \ref{fig:seds} shows the mid-IR SEDs for the four late-stage merger snapshots corresponding to the images in Figure \ref{fig:evol_with_images}, including the AGNx0 SEDs (for consistency with Figure \ref{fig:evol_with_images}, we use data from the same high-resolution simulation here, though the SEDs are very similar in the fiducial-resolution simulation). In snapshot ``a", the merging galaxies are just coming together for their final coalescence. The total AGN luminosity is low, $2\times 10^{43}$ erg s\inv, and the SED is overwhelmingly dominated by stellar emission. The AGN luminosity increases as the coalescence proceeds, both in absolute terms and relative to the total host luminosity. In snapshot ``c", the AGN contributes 40\% of the total luminosity, and the \wise\ \wonetwo\ color is $0.8$. At longer mid-IR wavelengths, prominent PAH emission and 9.7$\mu$m silicate absorption are apparent, reflecting the marginally dominant contribution of stellar emission to the total SED. At the peak of AGN activity (snapshot ``d"), where $L_{\rm AGN} = 7\times10^{45}$ erg s\inv, the AGN clearly dominates the total SED, which has a red slope from the near-IR through the 12$\mu$m  \wise\ $W3$ band. Here the AGN overwhelms most of the signatures of ongoing dusty star formation; the PAH emission and silicate absorption apparent in the AGNx0 simulation (and in the earlier fiducial snapshots) are much less prominent. The \wonetwo\ color is 1.3 at the AGN peak, which, given the 90\% contribution of the AGN to the total SED, is similar to the \wonetwo\ color of the intrinsic AGN SED template. 

The simulation shown in Figures \ref{fig:evol_with_images} \& \ref{fig:seds} is the high-resolution version of the A1A0 merger, with ten times higher mass resolution (and $10^{1/3}$ times higher spatial resolution) than the fiducial simulations. The higher spatial resolution reveals morphological features such as tidal tails, star forming regions, and dust-reddened nuclei in great detail in the images. The qualitative trends in $L_{\rm AGN}$, N$_{\rm H}$, and \wonetwo\ seen in Figure \ref{fig:evol_with_images} are generic to all of the major, gas-rich merger simulations; namely, we see a peak in activity soon after the galaxies' first pericentric passage, and a larger peak during final coalescence. The minor merger in our sample (A2A0, or simulation \#5) and the gas-poor mergers (\#6 \& 7) trigger significantly less star formation and AGN activity. In particular, the E1E0 gas-poor merger (\#7) never exceeds the $L_{\rm AGN} >10^{44}$ erg s\inv\ AGN threshold used in much of our analysis. Unless otherwise specified, the results below refer to the fiducial-resolution simulation suite; in Appendix \ref{ssec:resolution} we demonstrate the consistency of results between fiducial and high-resolution simulations.

\begin{figure}
\begin{center}
\includegraphics[width=0.49\textwidth]{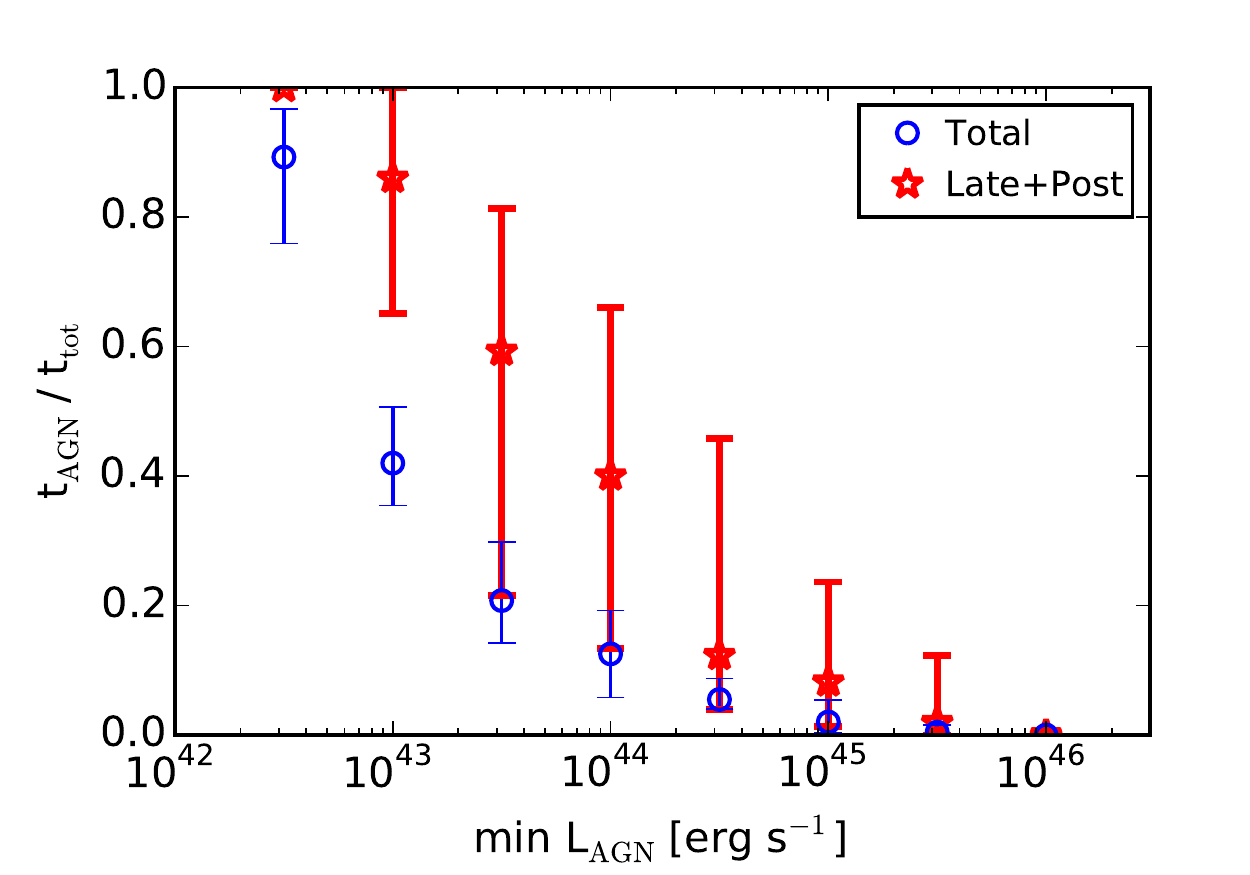}
\caption{The AGN duty cycle ($t_{\rm AGN}/t_{\rm tot}$) is shown for advanced mergers (Late+Post phases; red stars), versus the minimum bolometric luminosity used to define an AGN. For comparison, blue circles show the duty cycle for the entire evolution of the simulations. The data represent the median for all eight simulations in our suite, and error bars denote the inter-quartile range. AGN at all luminosities are much more likely to be found in the Late \& Post merger phases than in the Early phase, and AGN with luminosities of at least $10^{43}$ erg s\inv are active for $\ga 80\%$ of the Late+Post merger phases. \label{fig:agn_frac_allsim}}
\end{center}
\end{figure}

\subsection{(Mid-IR) AGN Lifetimes in Merging Galaxies}
\label{ssec:lifetimes}

For much our subsequent analysis, we divide each merger simulation into three phases: ``Early", ``Late", and ``Post-Merger", based on the BH separation. The ``Early" merger phase is defined as all times when the BH separation is $> 10$ kpc, the ``Late" phase is defined for BH separations $0 < a_{\rm sep} < 10$ kpc, and the ``Post-Merger" phase begins at the time of the BH merger. Unless otherwise noted, the BH separation $a_{\rm sep}$ refers to the {\em projected} separation, averaged over all viewing angles, and the Post-Merger phase is restricted to the first 100 Myr of post-BH-merger evolution. Because BH binary inspiral timescales are uncertain at sub-resolution scales (and are quite difficult to constrain observationally), we also combine the latter two phases (``Late+Post") for some of our analysis.

Using these definitions, we show in Figure \ref{fig:agn_frac_allsim} the fraction of time that an AGN is active in the Late$+$Post merger phases. We refer to this as the AGN duty cycle and show how it varies with the minimum observable AGN luminosity, and also how it compares with the AGN duty cycle for the entire merger simulation (including the Early phase). While the total duty cycle depends directly on the simulation duration, which is not a physical quantity, it is useful in relative terms when compared to the Late+Post stage duty cycle. As expected, AGN are on average much more active in the Late+Post merger phases. When low-luminosity AGN are included ($L_{\rm AGN} > 10^{43}$ erg s$^{-1}$), the BHs are typically active for $\ga 85\%$ of the Late+Post merger phase. For moderate-to-high luminosity AGN ($L_{\rm AGN} > 10^{44-45}$ erg s$^{-1}$), the AGN duty cycles are naturally shorter, and they depend more strongly on the merger parameters (particularly the initial gas fraction). The median AGN duty cycle at quasar luminosities ($L_{\rm AGN} > 10^{45}$ erg s\inv) is $\sim 7\%$ in the Late+Post merger phase, with a maximum of $\sim 50\%$ (in simulations 0 \& 1). These duty cycles correspond to quasar lifetimes ranging from 10 to 150 Myr in advanced mergers. 

\begin{figure}
\begin{center}
\includegraphics[width=0.49\textwidth]{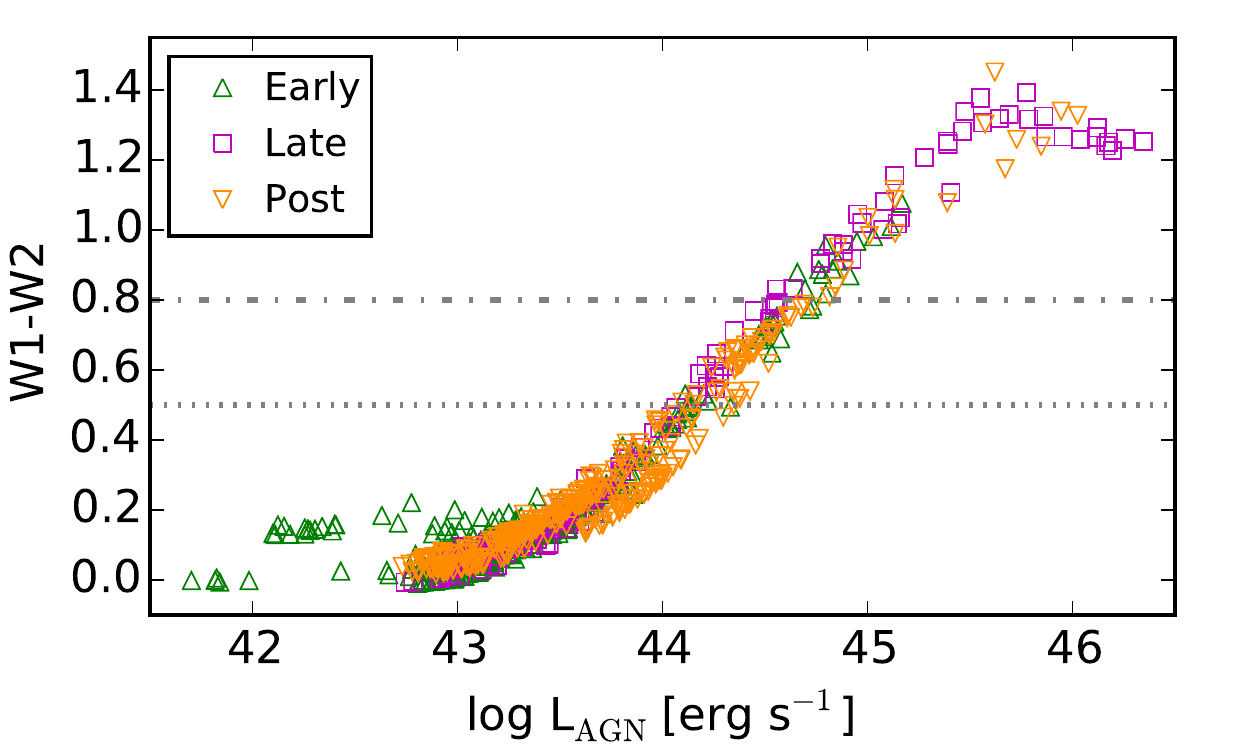}
\includegraphics[width=0.49\textwidth]{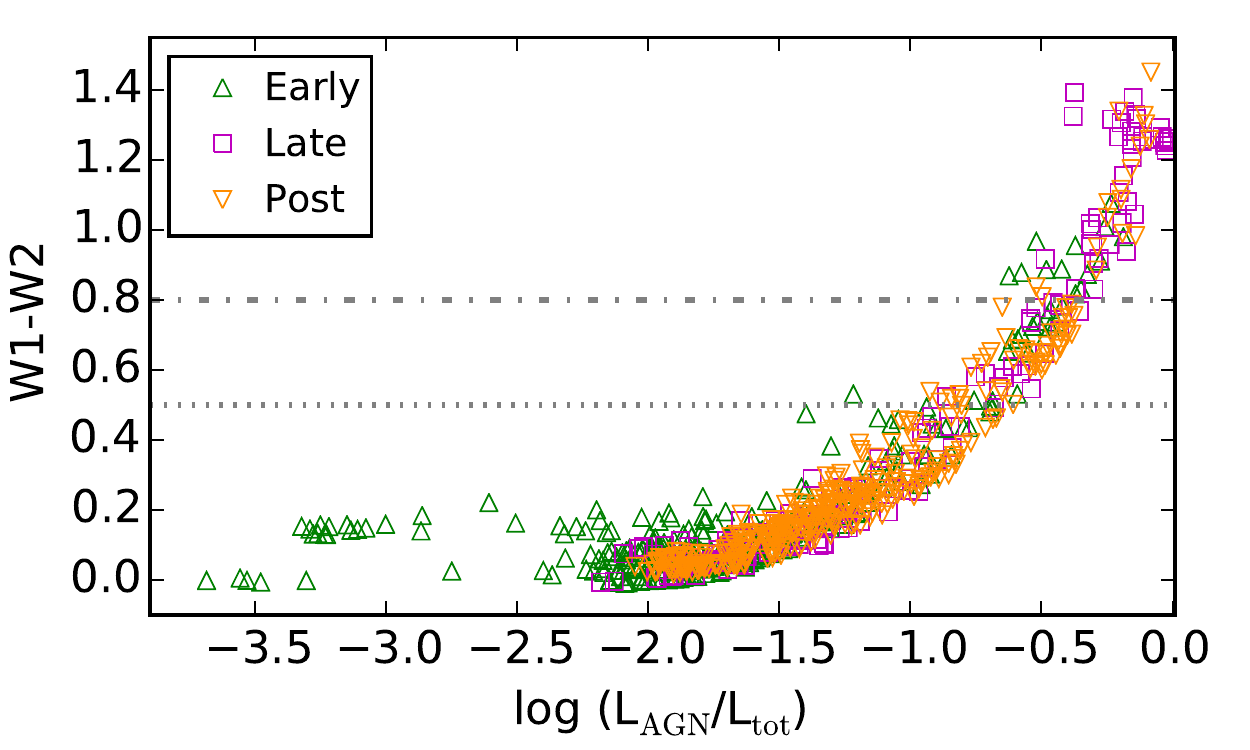}
\end{center}
\caption{In the top panel, the simulated \wise\ \wonetwo\ color is shown versus the AGN luminosity for all eight mergers in the simulation suite.  In the bottom panel, the AGN luminosity is shown as a fraction of the total bolometric luminosity in the host ($f_{\rm AGN} = L_{\rm AGN}/L_{\rm tot}$). Each point represents a single snapshot, where the \wise\ color is averaged over all sight lines. The green upward triangles, red squares, and orange downward triangles denote snapshots in the Early, Late, and Post-Merger stages, respectively. The gray dotted and dot-dashed lines denote the single-color cuts considered in this work (\wonetwo\ $> 0.5$ \& 0.8, respectively). The global \wise\ color is strongly correlated with the AGN contribution to the total host SED, with the reddest \wonetwo\ colors produced predominantly by luminous AGN in the Late and Post-Merger stages. \label{fig:w1w2_flbol_scatter}}
\end{figure}

We find a strong correlation between AGN luminosity and \wise\ mid-IR colors; as expected, the global \wonetwo\ color is reddest when the AGN dominates the total SED of the galaxy. To quantify this, we define the bolometric AGN fraction as $f_{\rm AGN} \equiv L_{\rm AGN}/L_{\rm tot}$, where $L_{\rm AGN}$ is the total {\em intrinsic} bolometric AGN luminosity (summed over both BHs if they have not yet merged) and $L_{\rm tot}$ is the total emergent bolometric luminosity of the system. Figure \ref{fig:w1w2_flbol_scatter} shows the correlation between $f_{\rm AGN}$ and the \wonetwo\ color for all eight fiducial simulations, with each point representing a different snapshot. The correlation has very little scatter across the range of merger environments and merger stages in our simulation suite, demonstrating clearly why mid-IR color selection is an effective AGN diagnostic. The data are also color-coded by merger stage, which illustrates that the highest luminosities (and reddest mid-IR colors) occur in the Late and Post-Merger phases. 

We see that $f_{\rm AGN} \ga 0.3$ is needed to achieve \wonetwo\ $> 0.8$; that is, the BH must contribute at least 30\% to the total bolometric luminosity to be selected as a mid-IR AGN via a single-color \wonetwo\ $> 0.8$ cut. Lower-luminosity AGN that would be missed by a \wonetwo\ $>0.8$ cut still contribute a non-negligible fraction of the total luminosity, $f_{\rm AGN} \ga 0.1$; these systems with $f_{\rm AGN} =$ 0.1 - 0.3 have bolometric luminosities in the range $L_{\rm AGN} =  6\times10^{43}$ - $6\times10^{44}$ erg s\inv.  Figure \ref{fig:w1w2_flbol_scatter} shows that a more lenient \wonetwo\ $>0.5$ color cut is sensitive to these moderate-luminosity AGN with $f_{\rm AGN} > 0.1$.

\begin{figure}
\begin{center}
\includegraphics[width=0.235\textwidth, trim={6.8cm 13.2cm 6.9cm 0.67cm},clip]{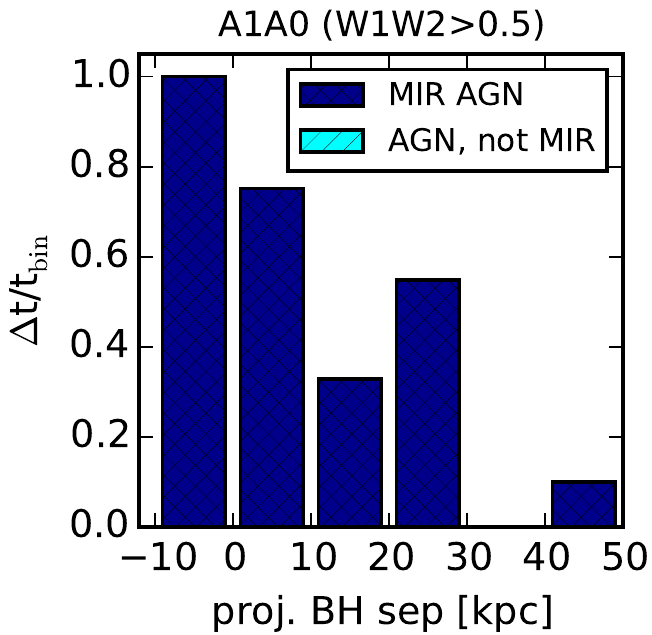}
\includegraphics[width=0.235\textwidth, trim={6.8cm 6.7cm 6.9cm 7.1cm},clip]{fig5.pdf}
\end{center}
\caption{For the A1A0 merger simulation, the AGN lifetime is shown versus projected separation, as a fraction of the total time in each separation bin (i.e., the AGN duty cycle). The total height of each bar (dark blue $+$ cyan) is the total (bolometric) AGN lifetime, where an AGN is defined as $L_{\rm AGN} > 10^{44}$ erg s\inv. The dark blue portion of the bar denotes the \wise-selected AGN duty cycle for \wonetwo\ $> 0.5$ ({\em left panel}) and for \wonetwo\ $> 0.8$ ({\em right panel}). The leftmost ``negative separation" bin denotes the Post-merger phase, capped at 100 Myr. AGN are preferentially triggered in the Late and Post-Merger stages (the two leftmost bins), and virtually all of this activity would be captured with a \wonetwo\ $>0.5$ color cut. With \wonetwo\ $>0.8$, however, these mid-IR AGN would be identified only about half of the time.  \label{fig:wise_lifetimes_example}}
\end{figure}

Figure \ref{fig:wise_lifetimes_example} shows the bolometric and mid-IR AGN duty cycles versus projected BH separation for an example simulation (A1A0), where we take $L_{\rm AGN} > 10^{44}$ erg s\inv. Here, the AGN duty cycle is calculated as a fraction of the total time spent in each separation bin, and mid-IR AGN duty cycles are shown for \wise\  \wonetwo $> 0.5$ and 0.8. As we saw in Figures \ref{fig:agn_frac_allsim} \& \ref{fig:w1w2_flbol_scatter}, AGN (including mid-IR AGN) are more likely to be found in the late stages of the merger. However, Figure \ref{fig:wise_lifetimes_example} makes clear that a more stringent \wise\ \wonetwo\ $>0.8$ selection criterion will miss most of the AGN lifetime in advanced mergers. In this example, a gas-rich, major merger where an AGN is active for all of the first 100 Myr of post-merger evolution, only 25\% of this AGN phase would be selected with \wonetwo\ $>0.8$, versus 100\% for \wonetwo\ $>0.5$. We find similar results if we compare these color cuts for even lower luminosity AGN, $L_{\rm AGN}>10^{43}$ erg s\inv. Conversely,  the most luminous AGN ($L_{\rm AGN}>10^{45}$ erg s\inv) are selected with high completeness with either single-color criterion, reflecting the fact that more stringent criteria will preferentially select the highest luminosity AGN.

\subsection{Reliability of Mid-IR AGN Selection in Mergers}
\label{ssec:reliability}

The primary concern with using a less stringent, more complete mid-IR selection criterion is the probable tradeoff in the reliability of AGN selection, with significant contamination from star-forming galaxies. Specifically, dust-enshrouded nuclear starbursts of sufficient intensity may mimic the red mid-IR SED slope characteristic of AGN-heated dust near its sublimation temperature ($\sim 1500$ K). (At high redshift, $z \ga 1$-1.5, the optical/near-IR peak of the stellar SED also begins to contaminate the mid-IR colors; we discuss results for $z>0$ in more detail below.) By comparing the set of fiducial \sunrise\ simulations with the corresponding AGNx0 simulations, in which the AGN luminosity is artificially set to zero, we can directly quantify the contamination of the mid-IR SED by star-formation-heated dust.

\begin{figure*}
\begin{center}
\includegraphics[width=0.9\textwidth,trim={0.1cm 0.1cm 0.1cm 0.6cm},clip]{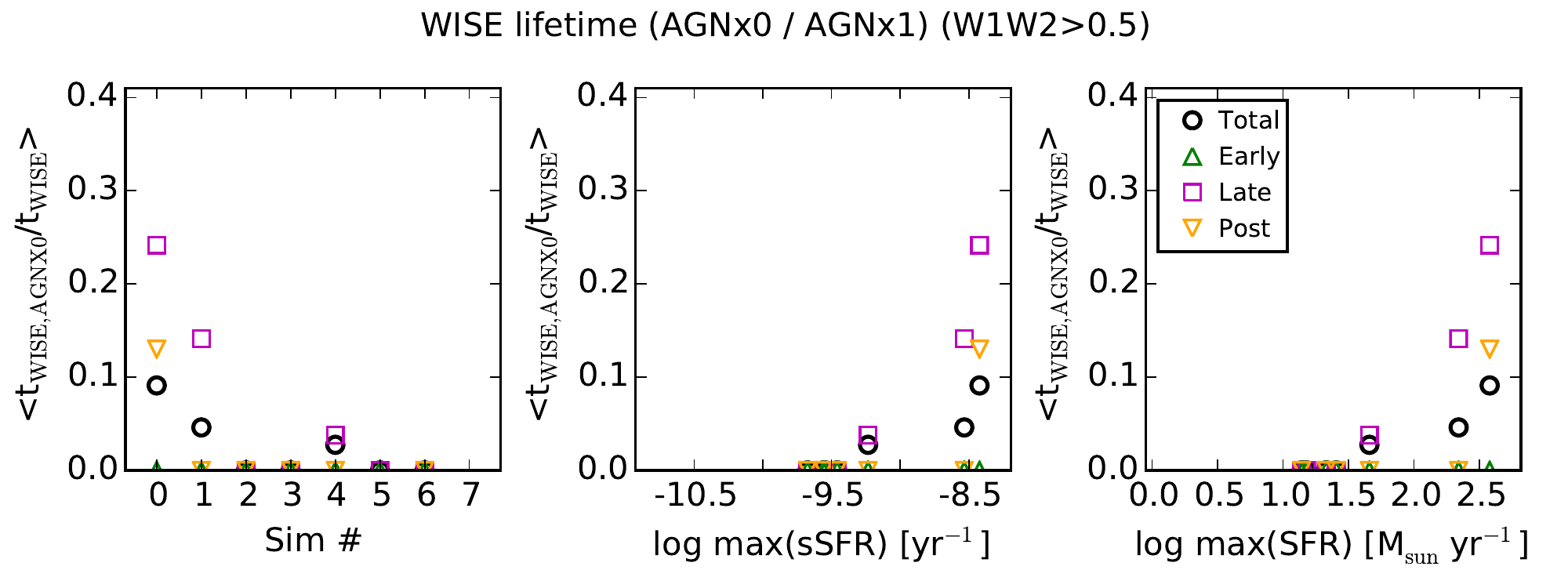}
\end{center}
\caption{For each simulation, and each merger phase, the ``contamination" fraction ($t_{\rm WISE,AGNx0}/t_{\rm WISE}$) of mid-IR colors by star formation in the AGNx0 simulations is shown. Specifically, this is the amount of time in each AGNx0 simulation for which \wise\ colors \wonetwo\ $>0.5$ are produced solely by star-formation heating, relative to the total \wise\ \wonetwo\ $>0.5$ lifetime in the corresponding fiducial simulation. All quantities are averaged over 7 viewing angles. The left panel shows results by simulation number as defined in Table \ref{table:merger_models}, and the middle and right panels show the contamination versus the maximum sSFR and SFR, respectively. 3/8 simulations have a nonzero lifetime with red \wise\ colors from star formation, occurring only during intense starbursts and constituting at most 15-25\% of the Late merger phase. When these starburst-induced red \wise\ phases occur in the AGNx0 simulation, the corresponding fiducial simulation {\em always} has a simultaneously-active AGN, so this is not ``true" contamination of the \wise\ AGN selection. \label{fig:sf_contam}}
\end{figure*}

\begin{figure*}
\begin{center}
\includegraphics[width=0.33\textwidth,trim={0.3cm 0.0cm 0.3cm 0.0cm},clip]{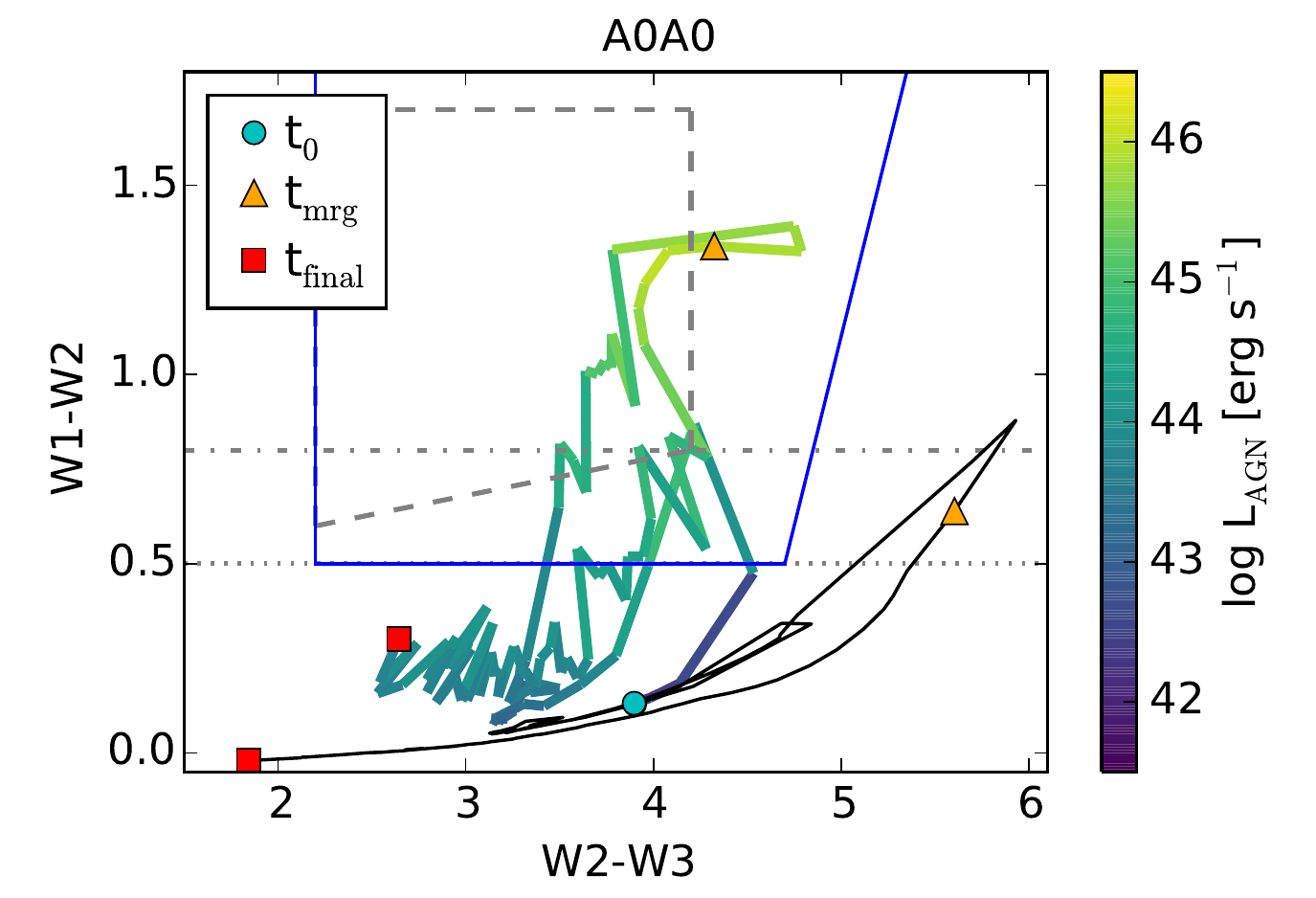}
\includegraphics[width=0.33\textwidth,trim={0.3cm 0.0cm 0.3cm 0.0cm},clip]{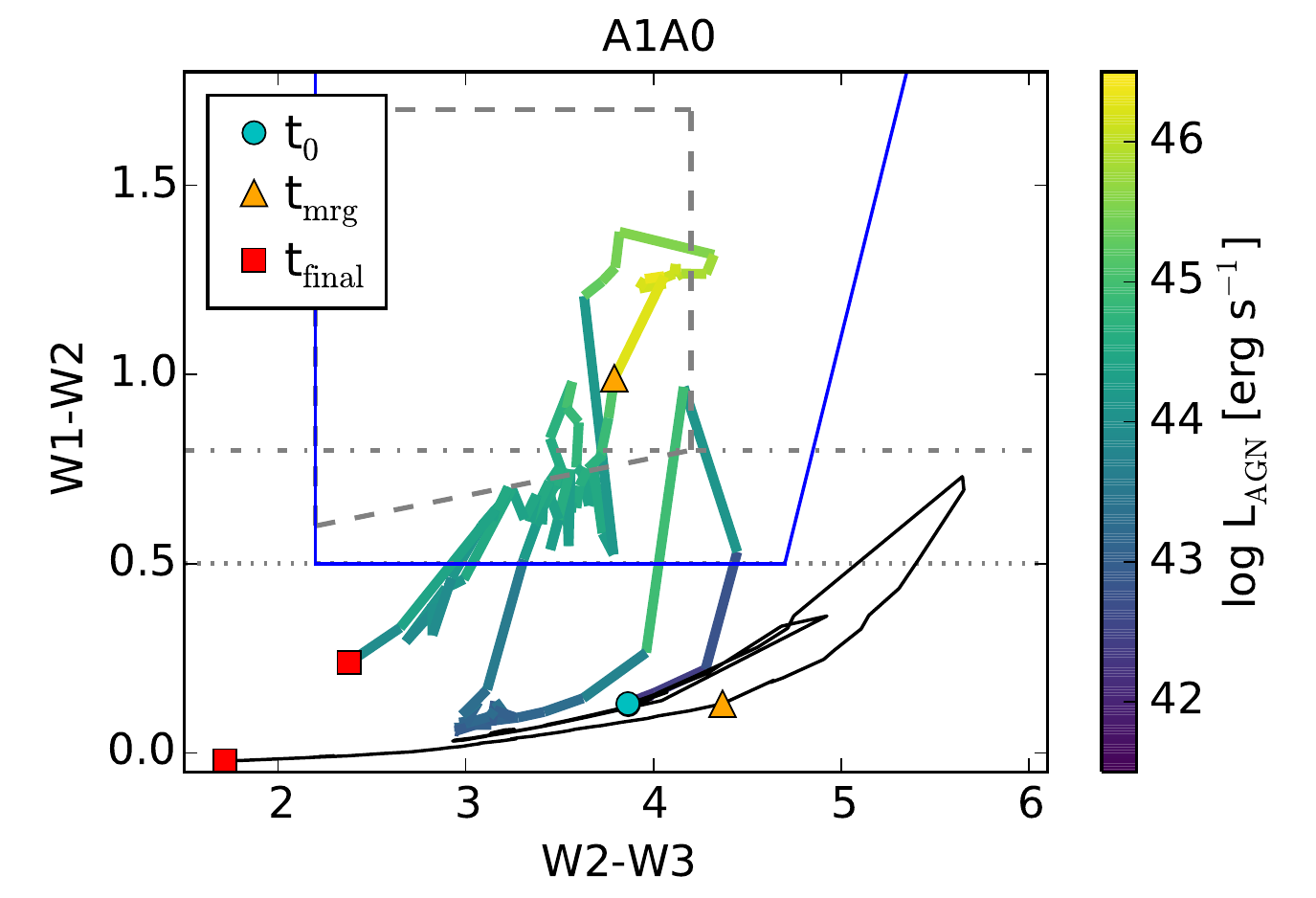}
\includegraphics[width=0.33\textwidth,trim={0.3cm 0.0cm 0.3cm 0.0cm},clip]{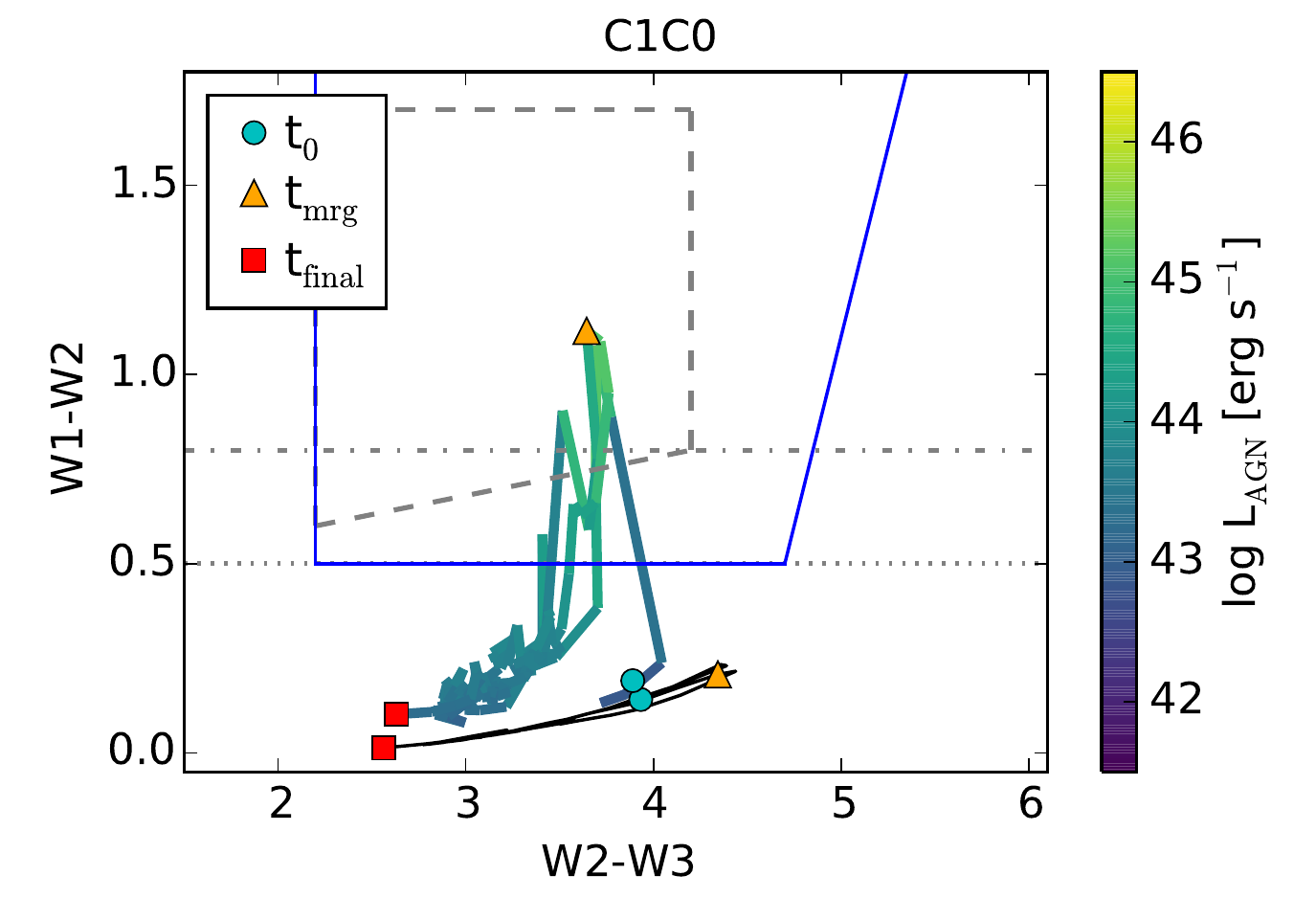}
\includegraphics[width=0.33\textwidth,trim={0.3cm 0.0cm 0.3cm 0.0cm},clip]{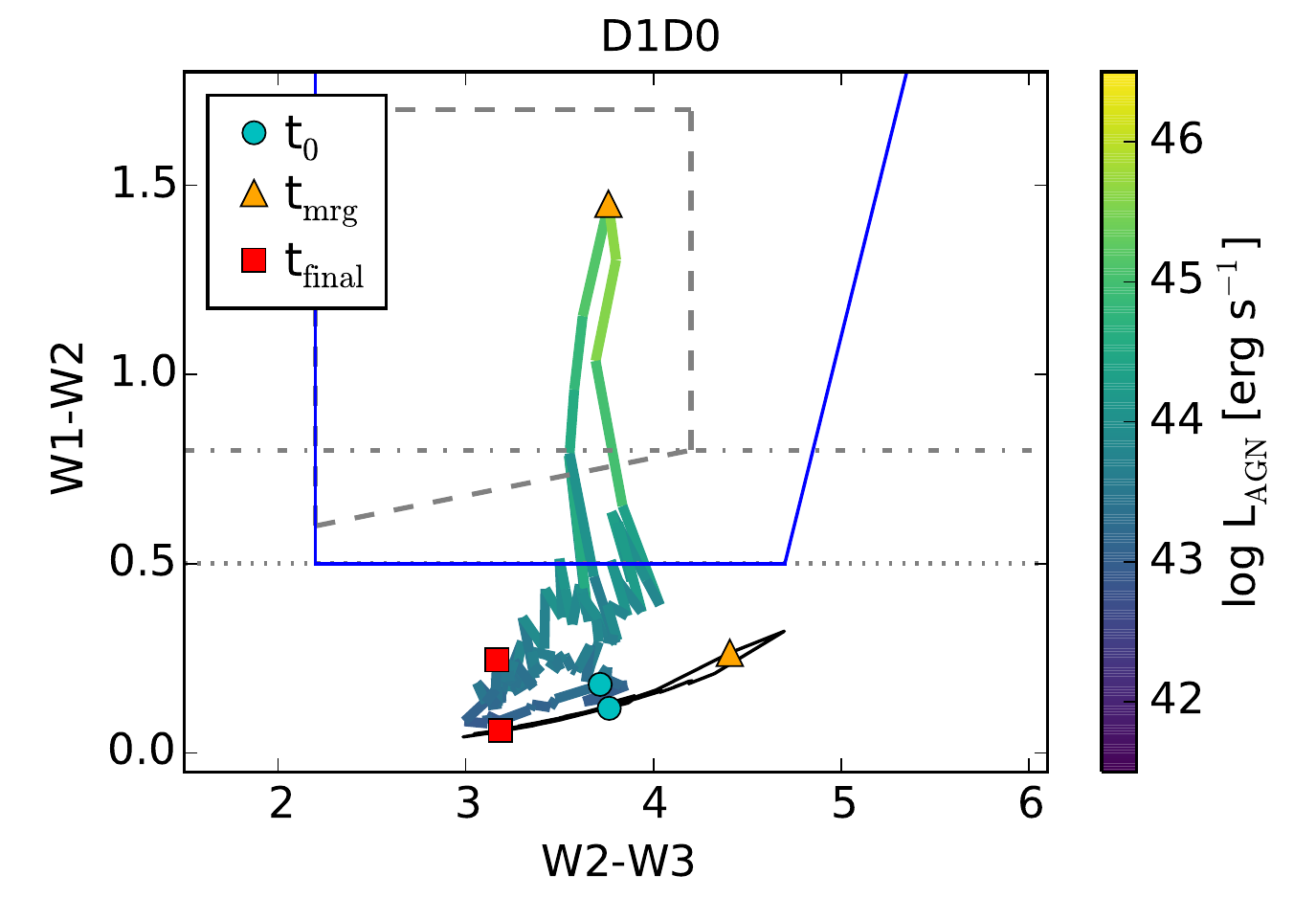}
\includegraphics[width=0.33\textwidth,trim={0.3cm 0.0cm 0.3cm 0.0cm},clip]{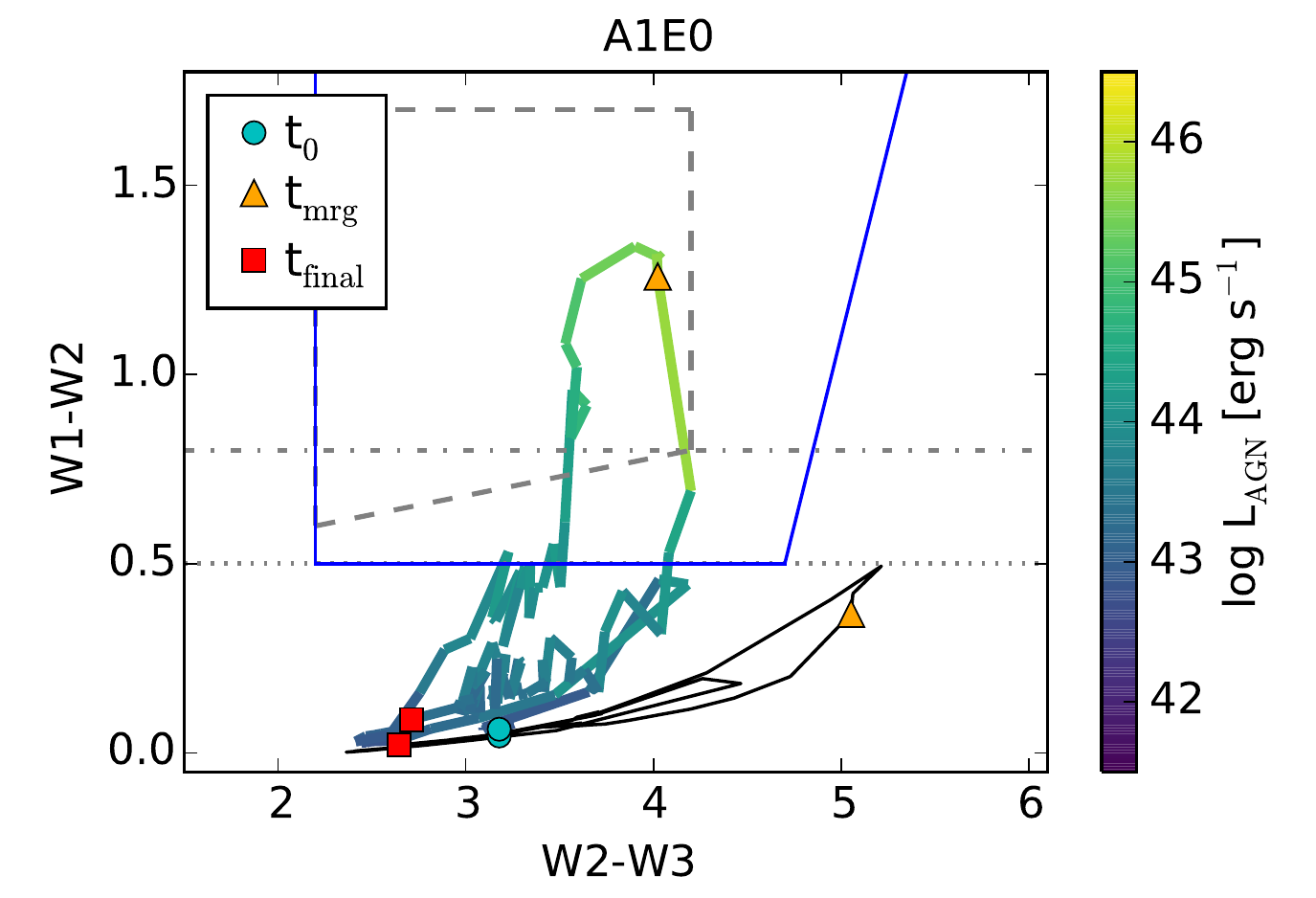}
\includegraphics[width=0.33\textwidth,trim={0.3cm 0.0cm 0.3cm 0.1cm},clip]{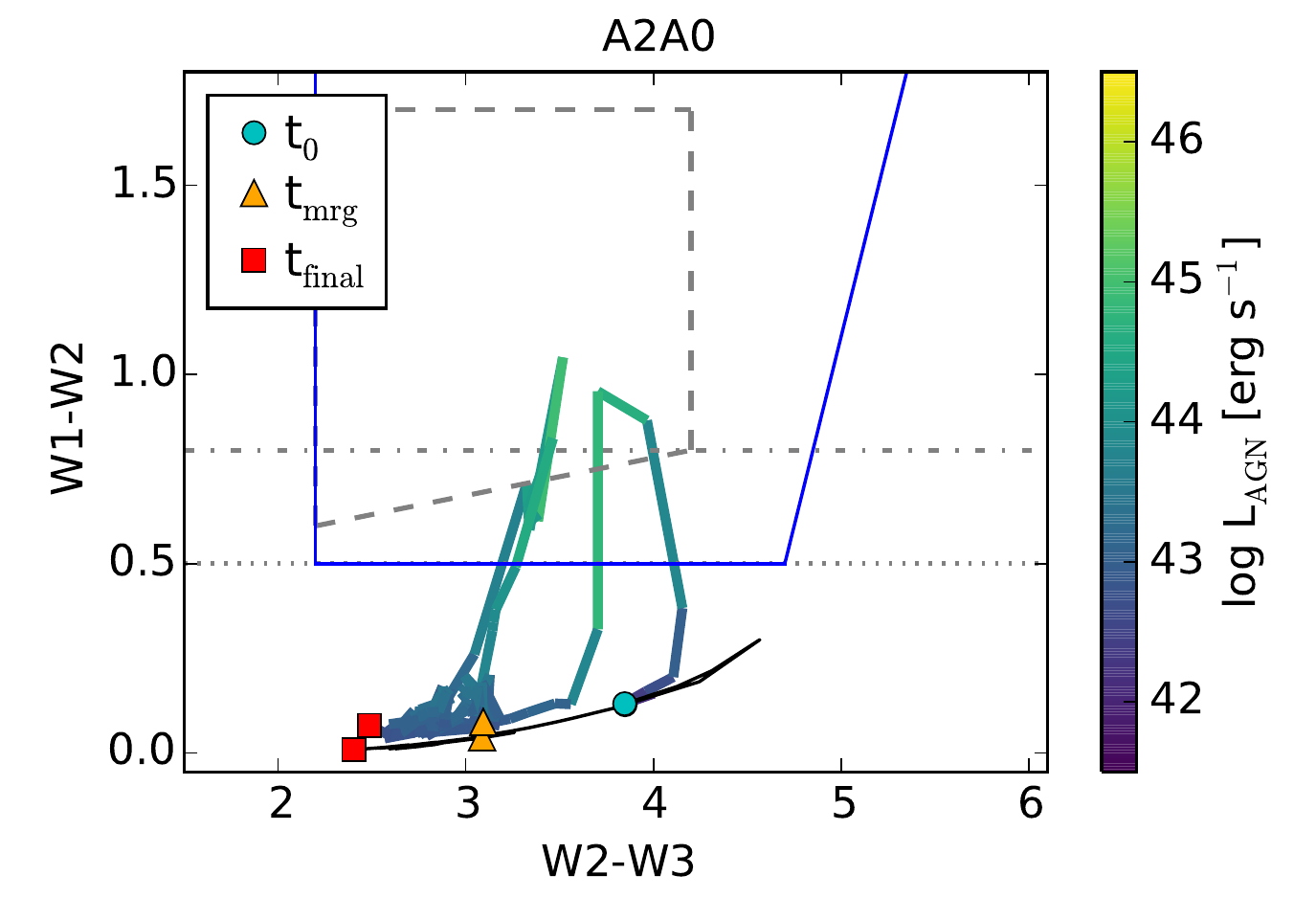}
\end{center}
\caption{The evolution of merging galaxies in \wise\ color-color space (\wonetwo\ versus \wtwothree) is shown for the gas-rich merger simulations (\# 0-5). The thick color curves show the evolution for the fiducial simulations, where the color scale denotes the bolometric AGN luminosity as shown in the color bar. The thin black curves show the evolution for the corresponding AGNx0 simulations (in which the AGN luminosity is artificially set to zero for the radiative transfer calculation). The cyan circles denote the initial snapshot, the orange triangles denote the time of BH merger, and the red squares denote the final snapshot at the end of the simulation. The gray dotted and dot-dashed lines show single-color \wonetwo\ selection cuts used in the literature and in this work. The gray dashed lines denote the two-color ``wedge" selection criteria of J11, and the blue solid lines denote the two-color cut motivated by this work (Equation \ref{eqn:mycut}, \S\ \ref{ssec:reliability}). The \wonetwo\ color typically peaks near the time of BH merger, and in the pure starburst (AGNx0) simulations it rarely exceeds $0.5$. The red \wtwothree\ colors produced in purely starburst systems are also clearly distinguishable from the AGN in color-color space. \label{fig:color-color}}
\end{figure*}

Figure \ref{fig:sf_contam} compares directly the mid-IR-selected lifetime of the fiducial versus the AGNx0 runs, thereby quantifying the contamination of mid-IR colors by star formation. We calculate the \wise-selected lifetime (for each merger stage) in each AGNx0 simulation relative to the \wise\ lifetime in the corresponding fiducial simulation ($t_{\rm WISE,AGNx0}/t_{\rm WISE}$). This measures the fraction of the mid-IR AGN lifetime when star formation alone would have been capable of producing the red \wise\ colors, if the AGN were not present. This fraction is shown In Figure \ref{fig:sf_contam} for \wonetwo\ $=0.5$, for each simulation by number, and also versus the maximum SFR and sSFR in each simulation. 

Only 3/8 merger simulations ever achieve \wonetwo\ $>0.5$ in the AGNx0 run, and the median contamination fraction ($t_{\rm WISE,AGNx0}/t_{\rm WISE}$) is zero for any single-color cut of \wonetwo\ $\geq 0.4$. One of these simulations (\#4, or A1E0) has $t_{\rm WISE,AGNx0}/t_{\rm WISE} \sim 4$\% in the Late merger phase; this corresponds to a single simulation snapshot. The other two simulations with nonzero contamination have very high maxmimum SFRs of 200-400 \msun\ yr\inv\ and sSFRs of $\sim 10^{-8.5}$ yr\inv. Empirical studies of galaxy pairs and galaxy mergers have shown that, at least at low to moderate redshift, such extreme starbursts are found almost exclusively in major mergers that are ULIRGs \citep[e.g.,][]{sanmir96,veille09}. These are short-lived and rare in the local Universe compared to less-extreme luminous IR galaxies (LIRGs, $L_{\rm IR} > 10^{11}$ L$_{\odot}$) with modest SFR enhancements \citep[][]{scudde12,elliso13a}. Even in these extreme starbursts, red \wonetwo\ $>0.5$ colors could only be produced by star formation alone for $\sim$ 15 - 25\% of the \wise\ lifetime in the Late merger stage, or 5 - 9\% of the total \wise\ lifetime. 

We now examine the evolution of the fiducial and AGNx0 simulations in \wise\ color-color space. Figure \ref{fig:color-color} shows the evolution of \wise\ \wonetwo\ versus \wtwothree\ color for four selected mergers, where the color and grayscale curves denote the fiducial and AGNx0 simulations, respectively. In the fiducial simulations, the \wonetwo\ color is low at the start of the simulations when the BH accretion rates are low, and then it rapidly reddens during peaks in AGN activity (generally after the galaxies' first pericentric passage and during their final coalescence; cf. Figure \ref{fig:evol_with_images}). Stellar and AGN feedback regulate these phases, and by the end of the simulation, the mid-IR colors are again dominated by stellar emission. The reddest \wonetwo\ colors are usually found during the final coalescence of the galaxies (for reference, the time of BH merger is marked with an orange triangle). Note, however, that the elevated \wonetwo\ colors during these AGN phases often fall outside the standard \wise\ selection cuts. At times this includes 
highly obscured ULIRGs ($L_{\rm IR} > 10^{12}$ L$_{\odot}$), with red \wtwothree\ colors that place them outside the two-color J11 wedge. In the A0A0 simulation, this reduces the completeness of the J11 two-color selection to 60\% in the Late merger phase. Finally, as in Figure \ref{fig:sf_contam}, Figure \ref{fig:color-color} makes clear that star formation alone is rarely  able to produce red, AGN-like \wonetwo\ colors in the host rest frame.

Motivated by our findings, we define a new two-color cut that leverages the higher completeness of a more lenient \wonetwo\ $>0.5$ cut for moderate-luminosity AGN. In addition to including many lower-luminosity AGN, these selection criteria will identify high-luminosity AGN in ULIRGs that coexist with starbursts and have red  \wonetwo\ and \wtwothree, while excluding pure starburst galaxies with very red \wtwothree\ (cf. Figure \ref{fig:color-color}). We also set a lower bound on \wtwothree\ of 2.2 mag, as in J11; none of our simulated galaxies have \wtwothree\ colors bluer than this at $z<1$. We propose the following criteria (in Vega magnitudes, as used throughout this work) for \wise\ color selection of AGN:
\begin{eqnarray}
\left[W1-W2\right] \: > \: 0.5 ,  \nonumber \\ 
\left[W2-W3\right] \: > \: 2.2 ,  \nonumber \\
{\rm and} \: \left[W1-W2\right] \: > \: 2.0 \times \left[W2-W3\right] - 8.9 
\label{eqn:mycut}
\end{eqnarray}
These criteria are shown in Figure \ref{fig:color-color} (blue lines) along with the other mid-IR selection criteria considered in this work. As Figure \ref{fig:wiseagn_frac_allsim} shows, this two-color cut yields very high completeness and reliability (median 100\%), for simulated merger-triggered AGN with $L_{\rm AGN} > 10^{44}$ erg s\inv.

As mentioned in \S~\ref{ssec:rt}, we have also repeated our analysis of mid-IR color selection for SEDs redshifted up to $z=1$ for all simulations, and up to $z=4$ for select simulation snapshots. For the redshifted SEDs, we find that contamination of the mid-IR AGN selection by starbursts is negligible for $z<1$ in our merger-triggered sample, for all one- and two-color selection criteria considered in this work except for \wonetwo\ $>0.5$ (which has at most 5-10\% contamination). At $z \sim$ 1 - 1.5, the optical/near-IR peak of stellar emission begins to shift into the {\em W1} and {\em W2} bands, greatly increasing contamination of the mid-IR AGN selection criteria \citep[see also the SED templates of][]{assef13}. As our study focuses on lower-redshift merging systems, in particular those with $z\la0.5$ where mid-IR AGN selection completeness is highest, the contamination of \wonetwo\ in high-redshift systems by star formation does not affect our conclusions based on rest-frame SEDs. Moreover, the \wtwothree\ color of star-forming galaxies becomes very blue ($<$ 2.2) at $2 \la z \la 4$, such that the two-color AGN criteria considered in this work do not suffer from such contamination in this redshift range.

Our discussion of the contamination of mid-IR AGN selection by star formation has thus far neglected a critical point: the quantity $t_{\rm WISE,AGNx0}/t_{\rm WISE}$ is not actually a true measure of the level of contamination. In reality, star formation and AGN fueling will often happen simultaneously, especially in gas-rich, late-stage major mergers. Thus, even when star-formation heating alone is {\em sufficient} to produce red \wonetwo\ colors, the more pertinent question is how often this happens {\em when an AGN is not simultaneously present}. In other words, we must calculate $t_{\rm WISE,AGNx0}/t_{\rm WISE}$ for {\em only} the AGNx0 snapshots in which the corresponding fiducial snapshot does not simultaneously have an AGN. We denote this new quantity as $t_{\rm WISE,SF}/t_{\rm WISE}$. For {\em every} merger simulation, we find $t_{\rm WISE,SF}/t_{\rm WISE} = 0$. That is, {\em all} of the merger-triggered starbursts that produce red \wise\ colors are coincident with merger triggered AGN. {\em Thus, in our simulations there is no ``true" contamination of the mid-IR AGN color selection, even for \wonetwo\ $>0.5$, or for the two-color criterion defined above.} 

Of course this doesn't mean that real samples of \wise-selected AGN have no contamination from starburst galaxies (e.g., \citealt{jarret11}; \citealt{stern12}; \citealt{mateos12}; \citealt{mateos13}; \citealt{assef13}).
As mentioned above, contamination increases sharply for $1 \la z \la 2$. We also do not model isolated galaxies, where secular processes may trigger less strongly-correlated star formation and AGN fueling, nor do we consider galaxies with BH masses that deviate substantially from the BH-bulge correlations. Even when star formation and AGN fueling are correlated, AGN are generally variable on very short timescales compared to the starburst lifetime (and the time resolution of the simulations); although IR variability timescales are longer than for optical or X-ray variability, this can decrease the amount of simultaneity. And clearly our finite simulation suite does not span the entire parameter space of merging systems. But for merger-triggered AGN at $z\la1$, we see that mid-IR color selection is very effective for the  selection criteria considered here, which are more lenient than those typically used in the literature.

\subsection{Completeness of Mid-IR AGN Selection in Mergers}
\label{ssec:completeness}

\begin{figure}
\begin{center}
\includegraphics[width=0.48\textwidth]{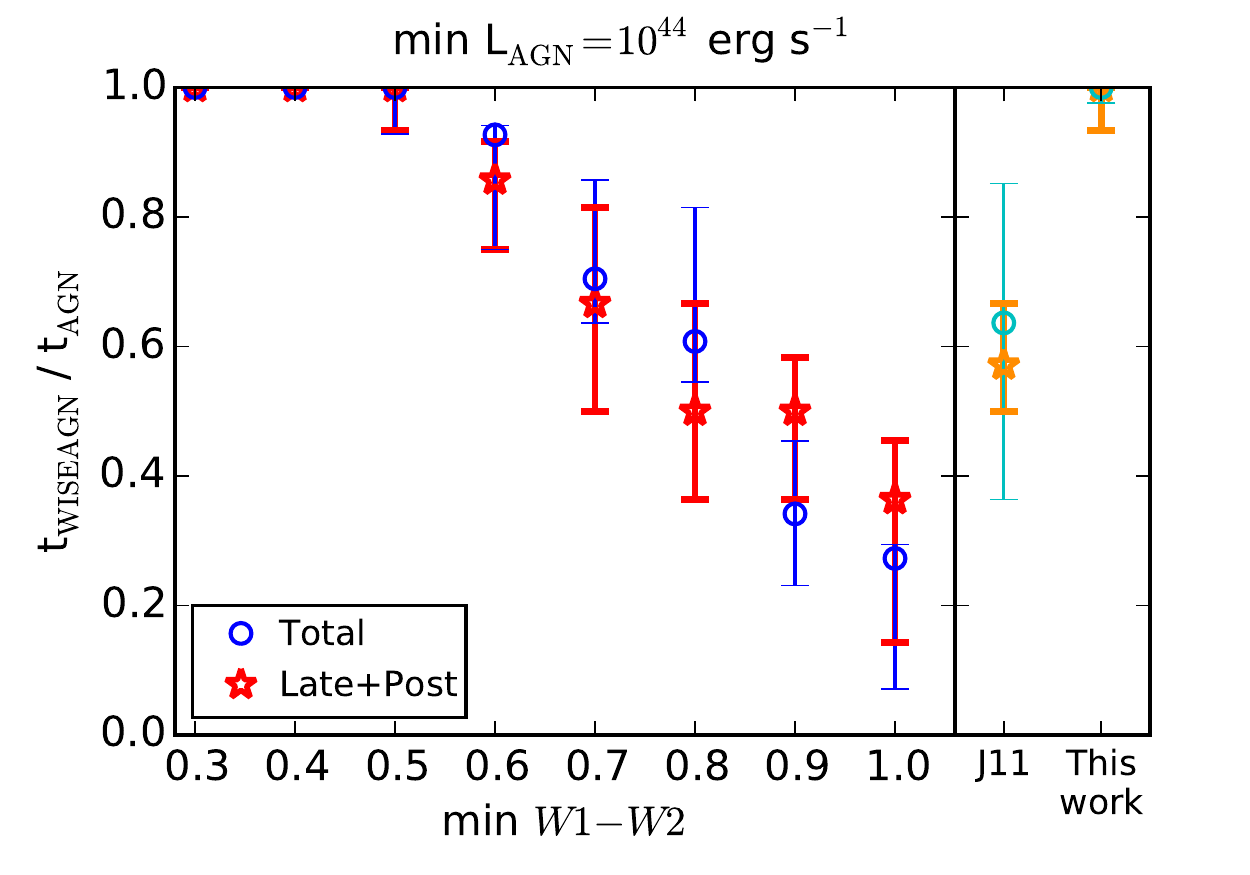}
\end{center}
\caption{The median mid-IR AGN selection completeness -- that is, the fraction of AGN that would be selected with a given mid-IR color cut -- is shown for all eight mergers in the simulation suite. The left side of the plots show selection completeness versus minimum \wonetwo\ for a single-color cut. The right side of the plots show the same quantities for the two-color cuts (\wonetwo\ and \wtwothree) of J11 and in this work (see \S\ \ref{ssec:reliability} for details). The blue and cyan circles denote the median \wise\ AGN fraction for the whole simulation, and the red and orange stars denote the median fraction for the Late+Post merger phases only. Error bars denote the inter-quartile range. For moderate-luminosity AGN ($L_{\rm bol} > 10^{44}$ erg s\inv), the completeness decreases steadily with stricter \wonetwo\ color cuts. The two-color cut proposed in this work (\S\ \ref{ssec:reliability}) has a median completeness of 100\% for moderate-to-high luminosity, merger-triggered AGN. \label{fig:wiseagn_frac_allsim}}
\end{figure}

In Figure \ref{fig:wiseagn_frac_allsim}, we quantify the dependence of AGN selection completeness on the mid-IR color criteria used, for moderate-to-high luminosity AGN ($L_{\rm AGN} > 10^{44-45}$ erg s\inv). The median fraction of AGN that would be selected with a given single-color \wonetwo\ cut is shown, for all simulations and for the Late+Post merger phase only. We also show the selection completeness for the two-color (\wonetwo\ \& \wtwothree) cut of J11, as well as the two-color cut proposed in this work (\S\ \ref{ssec:reliability}). 

At high AGN luminosities ($L_{\rm AGN} > 10^{45}$ erg s\inv), {\em all} mid-IR color cuts considered are essentially 100\% complete in merging systems. The J11 two-color criteria will miss some of the most luminous AGN in ULIRGs, where their red \wtwothree\ colors push them outside the selection region. The two-color cut defined in \S\ \ref{ssec:reliability} is chosen to include these luminous ULRIGs.

\begin{figure*}
\begin{center}
\includegraphics[width=0.7\textwidth,trim={0cm 7.3cm 0cm 0.3cm},clip]{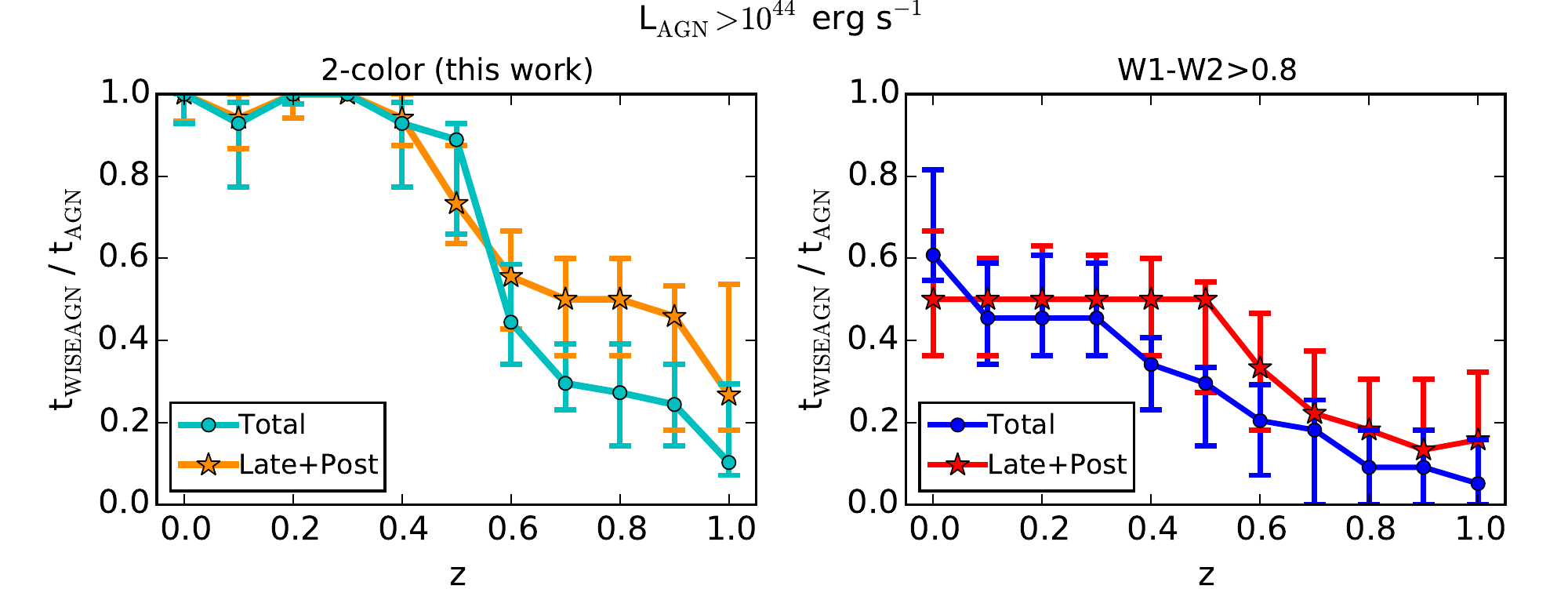}
\includegraphics[width=0.7\textwidth,trim={0cm 7.3cm 0cm 0.3cm},clip]{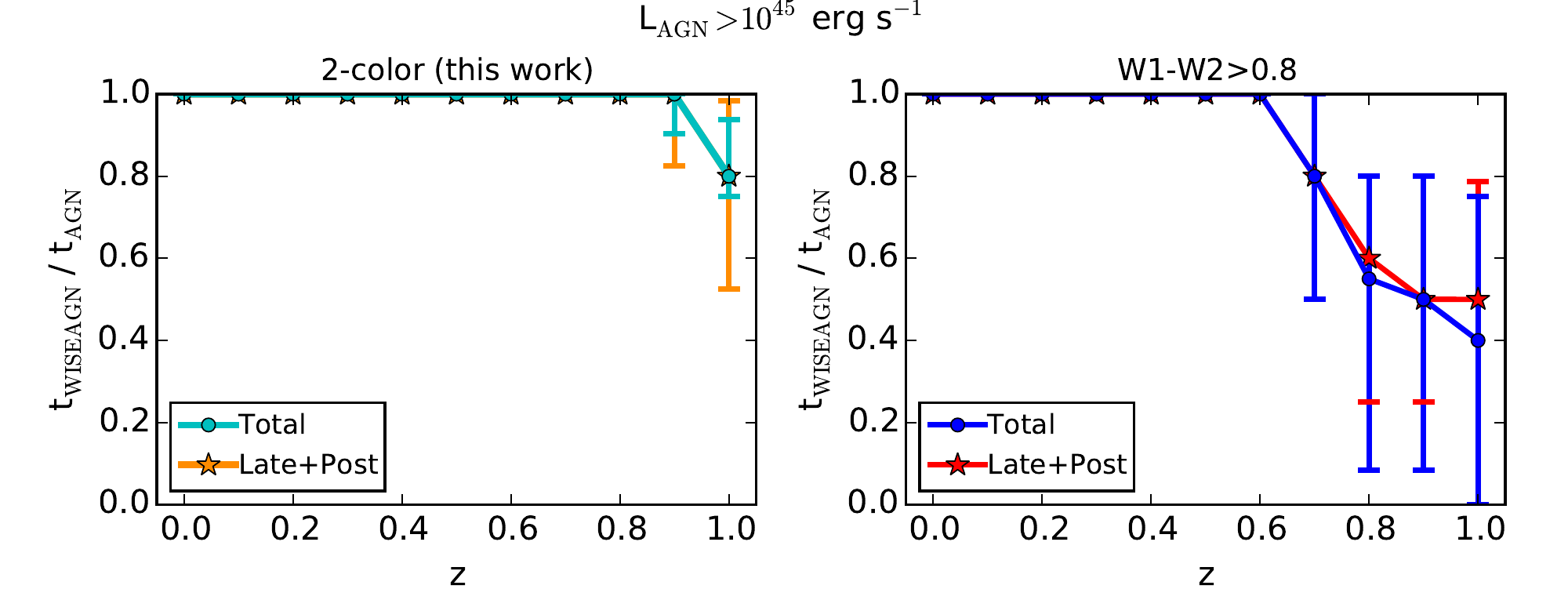}\end{center}
\caption{Mid-IR AGN selection completeness versus redshift is shown for the two-color cut proposed in this work (left panels) and for a single-color \wonetwo\ $>0.8$ cut (right panels). In all cases, a minimum bolometric luminosity is imposed ($L_{\rm AGN} > 10^{44}$ erg s\inv, top panels; $> 10^{45}$ erg s\inv, bottom panels). In addition, for $z>0$ we impose flux limits of $W1<18.50$ and $W2<17.11$, corresponding to the depth of the \wise\ survey in the NOAO Deep Wide-Field Survey Bo\"otes field at S/N $>3$ \citep{assef13}. In each panel, the completeness ($t_{\rm WISEAGN}$ / $t_{\rm AGN}$) is given for both the total merger simulation and for the Late+Post phase only. The two-color selection has high completeness ($\ga 80\%$) for moderate-luminosity AGN at $z \la 0.5$, much higher than for a stricter \wonetwo\ $>0.8$ cut. Even for high-luminosity AGN, the completeness of the \wonetwo\ $>0.8$ cut declines at $z\ga0.7$, while the two-color cut has $\sim 100\%$ completeness almost to $z \sim 1$.
\label{fig:completeness_vs_z}} 
\end{figure*}

For moderate-luminosity AGN ($L_{\rm AGN} > 10^{44}$ erg s\inv), mid-IR selection completeness is much more sensitive to the criteria used. Naturally, completeness decreases with increasingly color cuts. We see that $\sim$40-50\% of AGN with $L_{\rm AGN} > 10^{44}$ erg s\inv\ will be missed with a \wonetwo\ $>0.8$ criterion, {\em even in the late stages of gas-rich major mergers}. Similarly, the J11 two-color wedge is  $\sim 55-65\%$ complete for merger-triggered AGN at these luminosities. In these cases, the selection completeness for moderate-luminosity AGN varies substantially between different mergers, and between the Early, Late, and Post-merger phases. However, no clear systematic trends with merger stage or merger parameters are apparent; Figure \ref{fig:w1w2_flbol_scatter} suggests that this variation is driven in large part by the AGN luminosity relative to the host. Gas-rich mergers that produce luminous AGN will typically produce starbursts as well, thereby increasing the total luminosity of the host. Preferential nuclear obscuration during late-stage mergers also means that simultaneous star formation throughout the galaxy can more easily dilute the AGN signatures.

A more lenient AGN selection criterion of  \wonetwo\ $>0.5$ yields a median completeness of 100\% for moderate-luminosity AGN ($L_{\rm AGN} > 10^{44}$ erg s\inv) in our simulated merger sample. As Figure \ref{fig:wiseagn_frac_allsim} shows, this high completeness is independent of the merger parameters or merger stage. The two-color cut proposed in \S\ \ref{ssec:reliability} similarly has a median completeness of 100\% for these AGN luminosities. 

In addition to this analysis of the rest-frame simulated SEDs, we have calculated the completeness of mid-IR selection for SEDs redshifted up to $z=1.0$ (Figure \ref{fig:completeness_vs_z}). Here we have additionally imposed flux limits of $W1<18.50$ and $W2<17.11$, following \citet{assef13} as detailed in the figure caption. We find little difference in the mid-IR selection completeness for our simulated merger sample for $z\la0.5$. For moderate luminosity AGN, the median completeness drops below 90\% at $z=0.5$ for the two-color cut proposed in this work. The completeness for advanced mergers (in the Late+Post phases) drops less sharply with redshift, reflecting the redder colors of these obscured merging nuclei. Completeness for the \wonetwo\ $>0.8$ criterion is much lower overall, but follows a roughly similar redshift trend.

For luminous AGN ($> 10^{45}$ erg s\inv; lower panels of Figure \ref{fig:completeness_vs_z}), both the two-color cut and the \wonetwo\ $>0.8$ cut are 100\% complete out to $z=0.6$. At higher redshift, the completeness of the stricter single-color cut drops, while the completeness of the two-color cut proposed in this work remains at $\sim 100\%$ until $z=1$. At $z=1$, the completeness of the two-color cut is 80\%, versus 40-50\% for the \wonetwo\ $>0.8$ cut. Our results are therefore robust for moderate luminosity AGN at least to $z\sim0.5$ and for high luminosity AGN at $z<1$. Moreover, we see that the two-color cut is more effective at identifying merger-triggered AGN with high completeness than stricter cuts often used in the literature, even for high-luminosity AGN.

\subsection{Mid-IR Selection of Dual AGN}
\label{ssec:dual_agn}

The galaxy merger/AGN connection suggests that simultaneously-active {\em dual} AGN (AGN pairs with $\la 1$ - 10 kpc separations) are a natural consequence of merger-triggered BH fueling. Identification of such AGN pairs provides an unambiguous confirmation of a late-stage merger, acting as a sort of ``clock" that reveals the merger phase. They can also provide insight about the types of nuclear environments that are especially conducive to rapid BH fueling. Indeed, dual AGN activation becomes more likely as the merger progresses; \citet{elliso11} found a statistical excess of paired AGN that increases as pair separation decreases, and other studies have found similar trends \citep{koss12, satyap17}.
Although these spatially-resolvable AGN pairs are many orders of magnitude outside the gravitational wave (GW) dominated regime of BH inspiral, such objects are also the most readily accessible precursors to the massive BH binaries that will become GW sources. 

AGN pairs have been notoriously elusive, however; until recently only a handful were known, most of which were discovered serendipitously. Optical spectroscopic selection of AGN with double-peaked narrow-lines yielded a large sample of candidates, and at least 10\% of these do indeed appear to host dual AGN 
\citep[e.g.,][]{comerf12,liu12,comerf15}. However, the majority of double-peaked narrow-lines arise from outflows or other gas kinematics, and the fraction of time that dual AGN induce such features is intrinsically short 
\citep[e.g.,][]{blecha13b,muller15}.

Preferential obscuration of optical AGN signatures in mergers is another hindrance to identifying dual AGN via optical spectroscopy. In the hard X-ray selected AGN sample of \citet{koss10}, which reveals a much higher incidence of single AGN in mergers than optically-selected AGN samples, an even stronger excess of dual X-ray selected AGN is found. The dual AGN fraction is highest in the closest BH pairs, which are also the most luminous \citep{koss12}. Similarly, the strong enhancement in IR-selected AGN found in late-stage mergers \citep{satyap14} further suggests that obscured nuclear environments in late-stage mergers are likely sites for dual AGN. X-ray follow-up of these \wise-selected AGN in late-stage mergers has already revealed a high fraction of candidate duals \citep{satyap17,elliso17}, and near-IR coronal line emission detected in some objects provides further evidence for buried AGN in these systems \citep{satyap17}. 

\begin{figure}
\begin{center}
\includegraphics[width=0.23\textwidth, trim={6.8cm 6.8cm 7cm 7.1cm},clip]{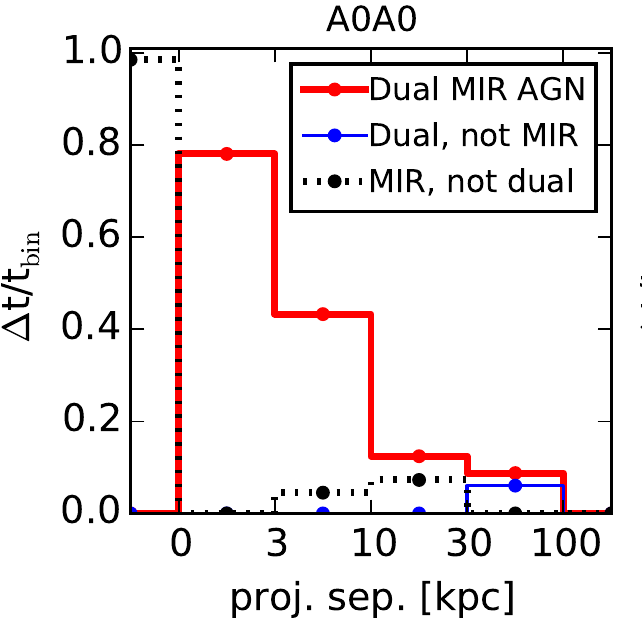}
\includegraphics[width=0.23\textwidth, trim={6.8cm 6.8cm 7cm 7.1cm},clip]{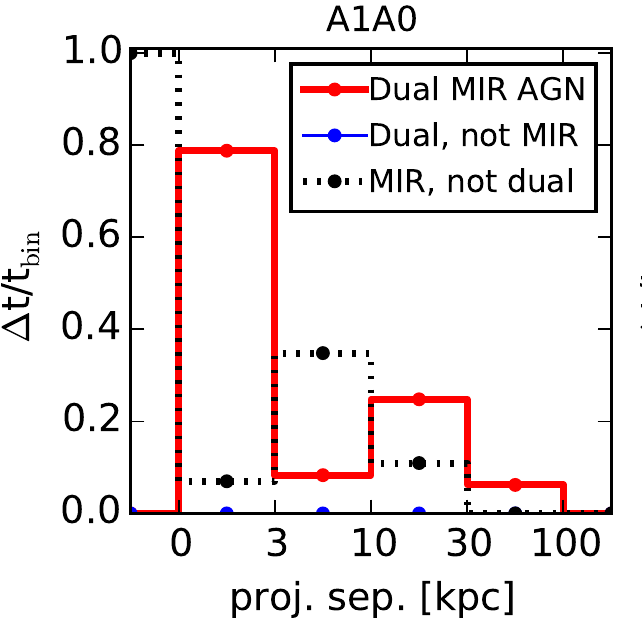}
\includegraphics[width=0.23\textwidth, trim={6.8cm 6.8cm 7cm 7.1cm},clip]{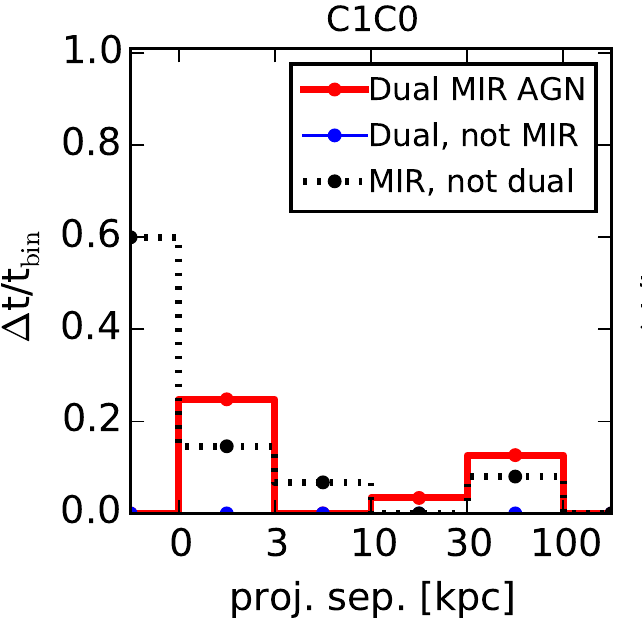}
\includegraphics[width=0.23\textwidth, trim={6.8cm 6.8cm 7cm 7.1cm},clip]{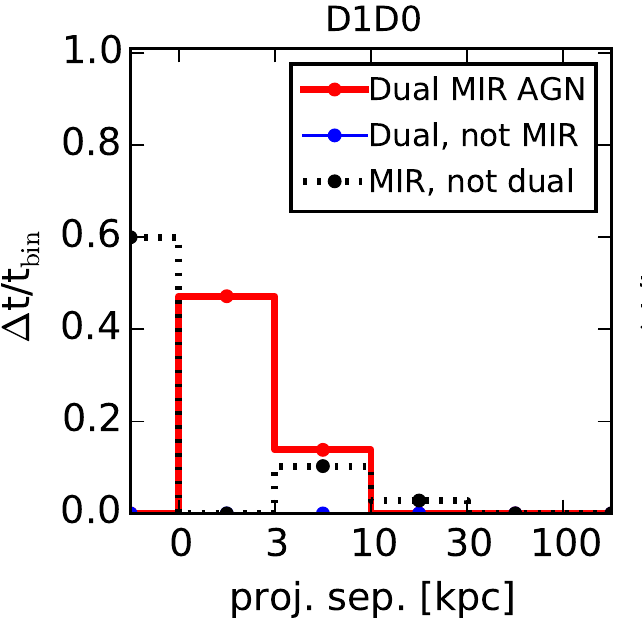}
\includegraphics[width=0.23\textwidth, trim={6.8cm 6.8cm 7cm 7.1cm},clip]{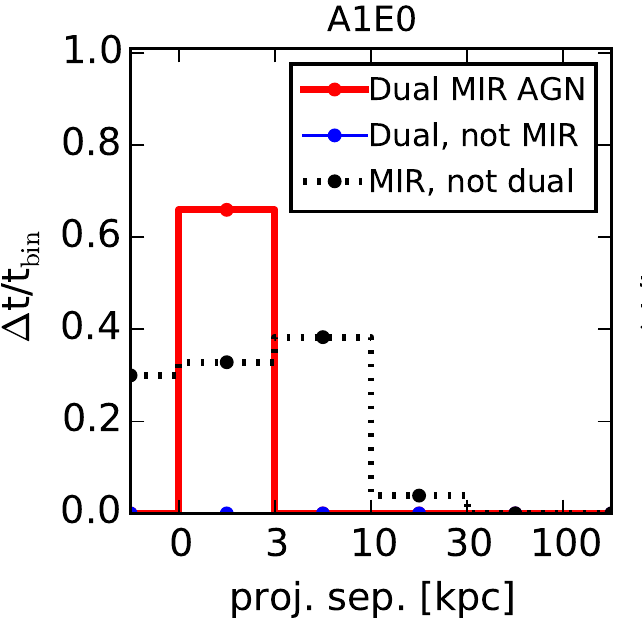}
\includegraphics[width=0.23\textwidth, trim={6.8cm 6.8cm 7cm 7.1cm},clip]{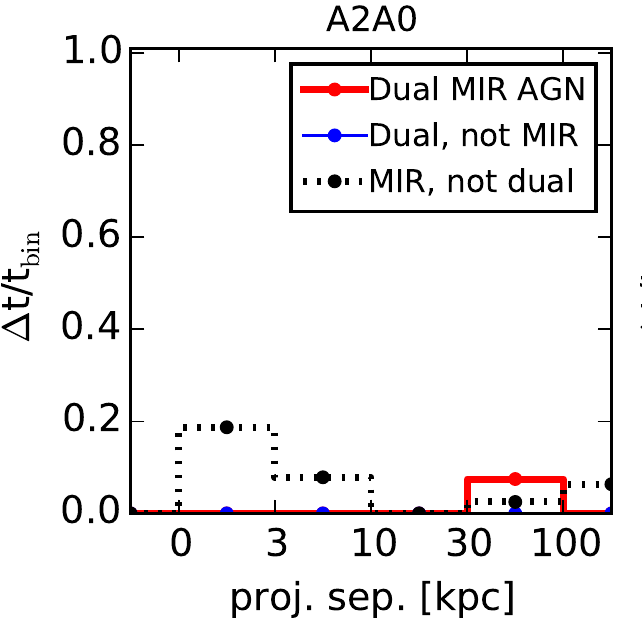}
\end{center}
\caption{For the six gas-rich merger simulations, the mid-IR dual AGN lifetime is shown versus projected separation, as a fraction of the total time in each separation bin (i.e., the dual AGN duty cycle). The leftmost ``negative" separation bin denotes the post-merger phase, capped at 100 Myr. Here, a ``dual AGN" is defined when each BH has a bolometric luminosity $> 10^{44}$ erg s\inv, and the system is considered a ``Dual MIR AGN" if the \wonetwo\ $>0.5$ criterion is simultaneously met. The thick red histogram denotes the \wise-selected dual AGN lifetime, while the dotted black histogram denotes \wise-selected AGN where only a {\em single} BH is active ($L_{\rm bol} > 10^{44}$ erg s\inv). The thin blue histogram denotes the phases where a dual AGN is present that would {\em not} be selected via \wise\ \wonetwo\ $>0.5$ colors; such a phase appears in only a single bin in the top-left plot. Thus, \wise\ color selection is extremely efficient at identifying dual AGN hosts, particularly at small separations where many such duals may still be unresolved. 
\label{fig:wise_dual_projsep_w05}}
\end{figure}

Using our merger simulations, we quantify the expected fraction of dual AGN in each merger stage, in terms of both bolometric AGN luminosity and mid-IR color selection. Figure \ref{fig:wise_dual_projsep_w05} shows the fraction of time at each separation when both BHs are simultaneously active, separated by whether or not the dual AGN would be selected via a \wonetwo\ $>0.5$ \wise\ color cut. Also shown is the time for which only a single \wise\ AGN is present. We see that dual AGN duty cycle is strongly peaked at the smallest BH separations, $< 3$ kpc. Thus, even though dual BHs spend less time at small separations than at larger separations, late-stage mergers are much more likely to contain dual AGN.

\begin{figure}
\begin{center}
\includegraphics[width=0.45\textwidth]{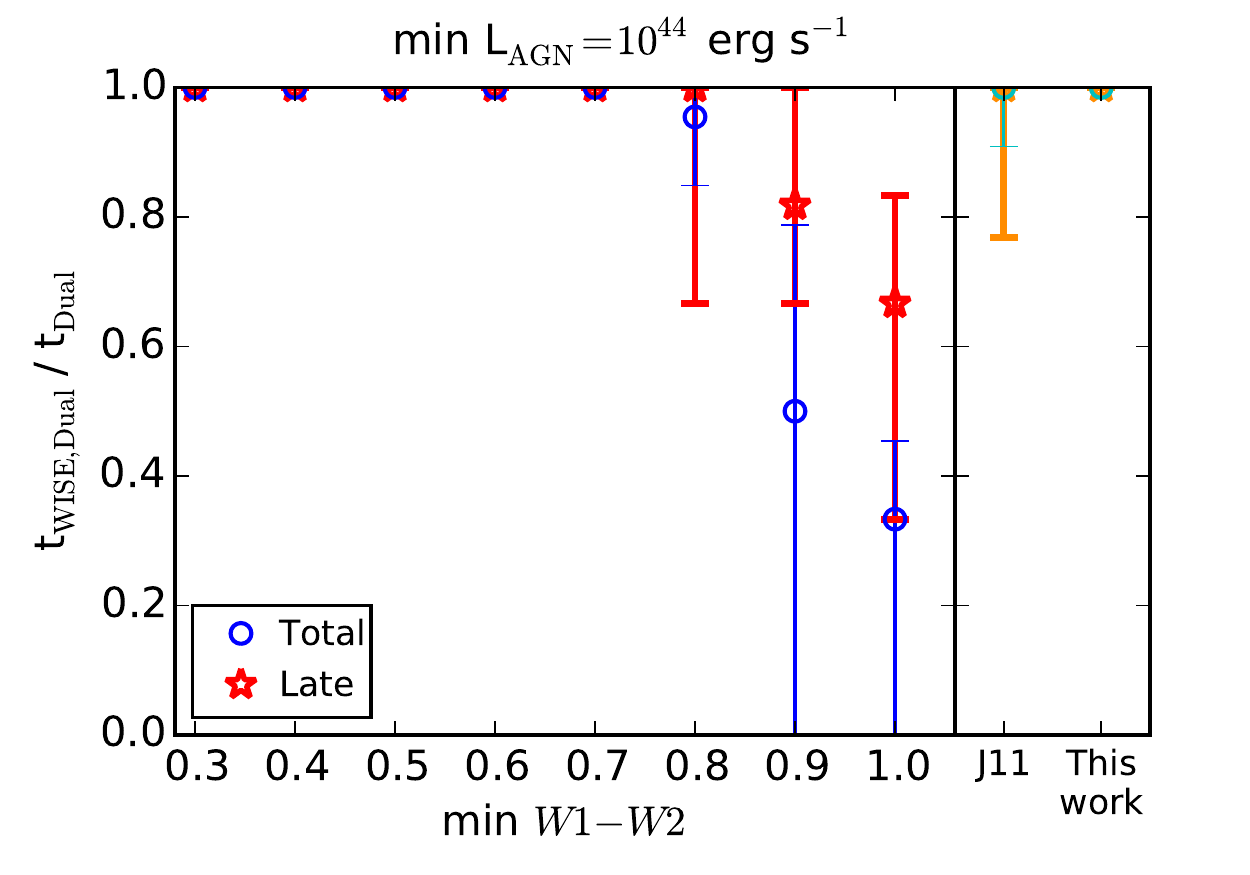}
\end{center}
\caption{The mid-IR selection completeness for {dual} AGN (pairs of simultaneously-active AGN) is shown in a similar manner as the single-AGN completeness in Figure \ref{fig:wiseagn_frac_allsim}. Specifically, the  \wise-selected dual AGN lifetime is plotted as a fraction of the total dual AGN lifetime. Blue circles and red stars denote the median fraction for the total merger and the Late phase, respectively, versus the \wonetwo\ single-color cut assumed. The cyan circles and orange stars similarly show $t_{\rm WISE,Dual}/t_{\rm Dual}$ for the two-color cuts defined in J11 and in this work. Dual AGN are defined as systems where each BH exceeds the minimum luminosity threshold $L_{\rm AGN} > 10^{44}$ erg s\inv. Mid-IR selection is even more effective at identifying dual AGN than single AGN, with a median completeness of $\sim$ 100\% for all the selection criteria considered in this work. \label{fig:wise_dual_frac}}
\end{figure}

Also notable is the fact that, for $L_{\rm AGN} > 10^{44}$ erg s\inv, mid-IR AGN selection identifies virtually {\em all} dual AGN systems. Figure  \ref{fig:wise_dual_frac} shows the fraction of the dual AGN lifetime for which the system would be selected as a \wise\ AGN, for each merger phase. The median completeness of dual AGN selection is near 100\% for {\em all} single- and two-color mid-IR selection criteria considered in this work. For the more stringent cuts in particular (\wonetwo\ $>0.8$ and the J11 two-color cut), the dual selection completeness is significantly higher than the single-AGN completeness (Figure \ref{fig:wiseagn_frac_allsim}), which is typically $\sim 50$-60\%. The higher completeness for \wise\ selection of dual AGN largely owes to the fact that the conditions conducive to fueling both BHs simultaneously (i.e., a central, dense reservoir of cold gas) are also conducive to fueling luminous, obscured AGN, which produce red mid-IR colors. It is perhaps unsurprising, then, that mid-IR color selection is so effective at identifying merging hosts of dual AGN.

\begin{figure}
\begin{center}
\includegraphics[width=0.45\textwidth]{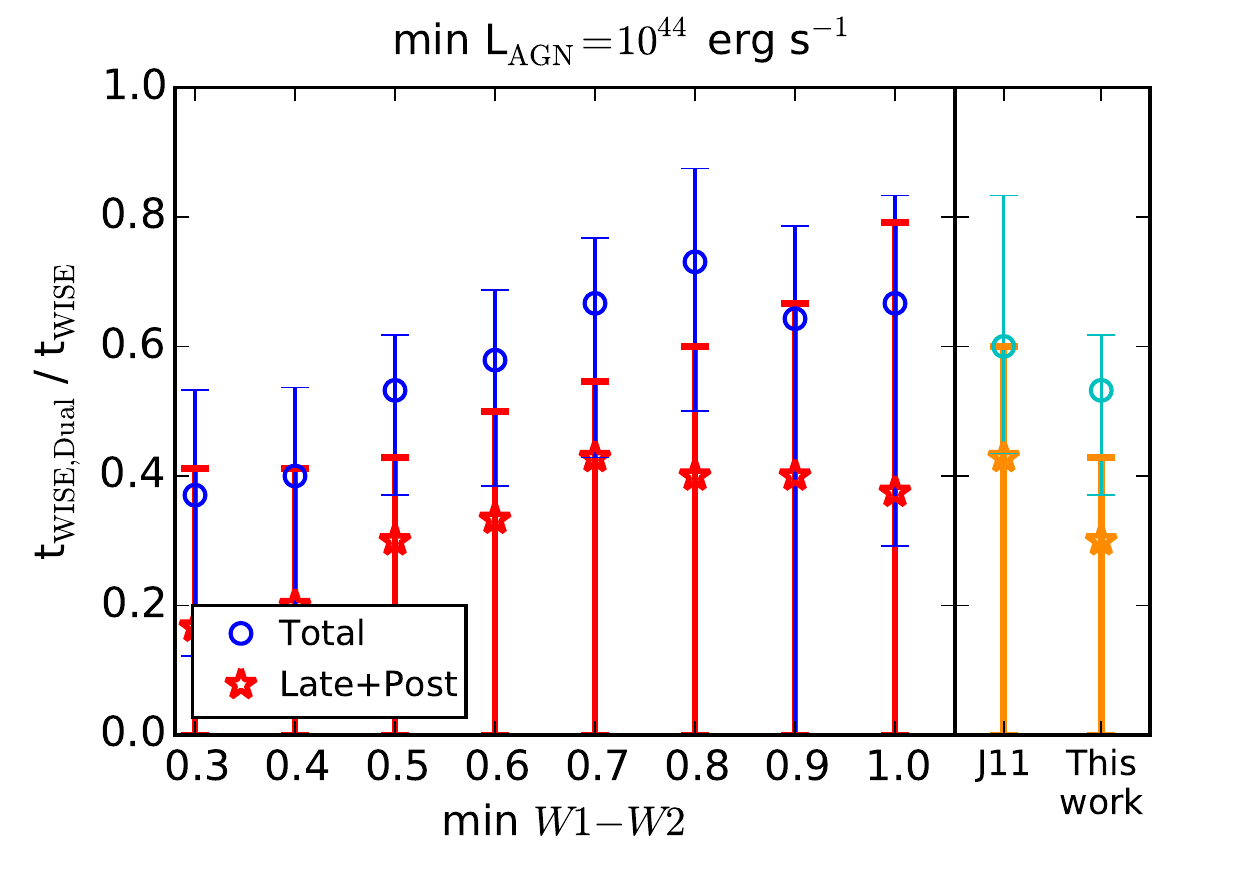}
\end{center}
\caption{The fraction of the {\em total} mid-IR lifetime for which a dual AGN is present ($t_{\rm WISE,Dual}/t_{\rm WISE}$) is shown for various mid-IR selection criteria. Blue (cyan) circles denote this fraction for the total merger (for single- \& two-color \wise\ cuts, respectively), and red (orange) stars similarly denote the fraction for the Late+Post merger phase. Note that the dual AGN lifetime is zero by definition in the Post-merger phase. Although dual mid-IR AGN are efficiently triggered advanced mergers (Figure \ref{fig:wise_dual_projsep_w05}), the Late merger phase is intrinsically short. The fraction of \wise\ AGN containing duals varies greatly between mergers, particularly in the Late phase, and the short lifetime of the Late phase yields a somewhat higher probability of finding dual AGN at larger separations. Nonetheless, we find that most merger-triggered mid-IR AGN, and $\sim$ 30-40\% of mid-IR AGN in advanced mergers, are actually dual AGN with $L_{\rm AGN} > 10^{44}$ erg s\inv. 
\label{fig:wise_dual_to_wise_allsim}}
\end{figure}

For observed samples of \wise-selected AGN, we wish to know not only the {\em completeness} of dual AGN selection, ($t_{\rm WISE,dual}/t_{\rm dual}$), but also the fraction of {\em all} \wise\ AGN that contain duals ($t_{\rm WISE,dual}/t_{\rm WISE}$). Figure \ref{fig:wise_dual_to_wise_allsim} shows this fraction for our simulated sample of merger-triggered AGN, for the total merger and for the Late+Post merger phases (note that the dual AGN fraction in the Post-merger phase is zero by definition). The fraction $t_{\rm WISE,dual}/t_{\rm WISE}$ varies greatly between merger simulations, particularly for the advanced mergers. In the Late phase, 3/8 simulations (the minor merger A2A0 and the gas-poor simulations) have no dual AGN activity and little to no single AGN activity for $L_{\rm AGN} >10^{44}$ erg s\inv. 
Moreover, although a high fraction of advanced mergers contain dual mid-IR AGN (Figure \ref{fig:wise_dual_projsep_w05}), the Late merger phase prior to BH merger is intrinsically short. Overall, we see that dual mid-IR AGN are actually somewhat more likely to be found at larger separations ($\sim$ 10-30 kpc). For the mid-IR selection criteria we consider, we find that a majority (median fraction $\sim$ 55-75\%) of merger-triggered mid-IR AGN contain an AGN pair. And even for the Late+Post merger phases, where a dual AGN can exist only in the Late phase, we still find that a significant fraction ($\sim$ 30-40\%) of mid-IR AGN are expected to be duals.

The fraction of mid-IR AGN in duals is highest for the more stringent mid-IR selection criteria (\wonetwo\ $>0.8$ and the J11 two-color cut), typically $\sim$ 60-85\% of the total \wise\ lifetime. The stricter color cut yields a slightly higher fraction of dual AGN because of the correlations between AGN luminosity and mid-IR colors, and between AGN luminosity and dual AGN triggering. In other words, these criteria are sensitive to the most luminous AGN, and the same merger-driven dynamics that produce these luminous AGN will increase the probability that both BHs are simultaneously active. 

We reiterate that the finite time and spatial resolution of the simulations limits their sensitivity to AGN variability. Uncorrelated, large-amplitude stochastic variability in each AGN would reduce the dual active lifetime compared to our simulations. This is unlikely to be as important in the IR, where AGN are less variable and timescales are longer, but it could affect the fraction of \wise-selected AGN with dual X-ray detected nuclei, for example. Nonetheless, our results indicate that a significant fraction of merger-triggered mid-IR AGN should contain dual active nuclei, and the early success of follow-up programs in identifying candidates supports this conclusion \citep{satyap17,elliso17}.

Finally, we saw in Figure \ref{fig:wise_dual_projsep_w05} that \wise-selected dual AGN are most likely to be found at the smallest BH separations. Dual mid-IR AGN (with $L_{\rm AGN} > 10^{44}$ erg s\inv) are present in up to 80\% of mergers with projected separations $< 3$ kpc. This demonstrates that many  dual AGN in \wise-selected samples are likely still unresolved.  \wise-selected AGN in late-stage mergers are therefore promising targets for further follow-up studies with high-resolution X-ray, optical, or IR imaging and resolved spectroscopy.

\subsection{Merger-driven Obscuration of AGN}
\label{ssec:obscuration}

Our finding that mid-IR AGN selection is highly effective in advanced mergers supports the idea that some AGN obscuration is environmentally-driven, rather than depending primarily on viewing angle as in standard AGN unification scenarios. Merger-driven dynamics may create significant obscuration on galactic scales,  increase the dust covering fraction of the nuclear ``torus" on smaller scales, or both. While we do not resolve the AGN torus scale in the simulations, we can quantify the relative amount of galactic-scale obscuration throughout the merger.

\begin{figure*}
\begin{center}
\includegraphics[width=0.7\textwidth]{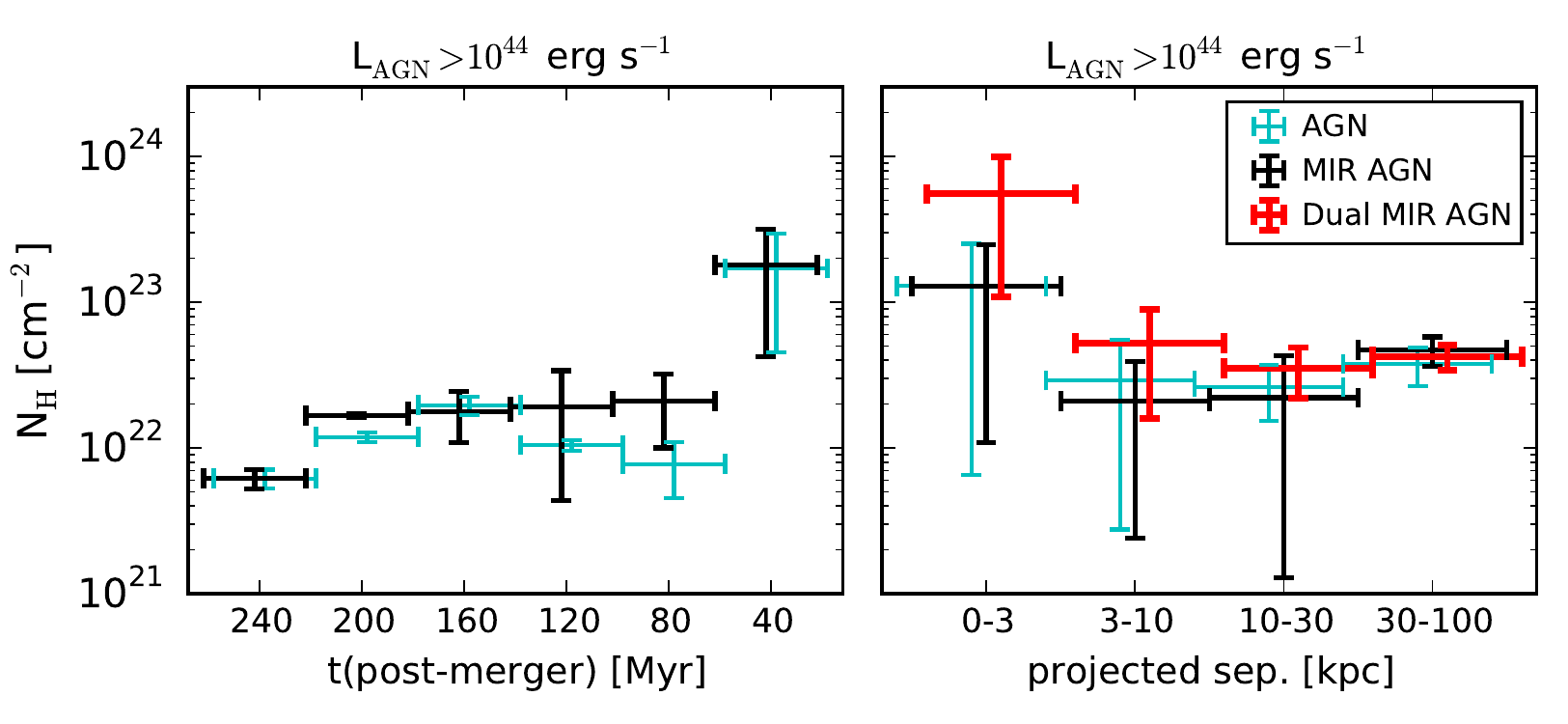}
\end{center}
\caption{The total gas column density (\nh) along the AGN line of sight is shown versus merger stage; this provides an upper limit on the amount of galactic-scale obscuration. Prior to the BH merger, the merger evolution is binned by projected BH separation, decreasing from right to left (following the convention used in the literature and throughout this work). After the BH merger, the evolution is binned by the time elapsed since the BH merger, increasing from right to left (such that time evolution can be followed from right to left across both pre- and post-merger panels). \nh\ is shown for AGN with bolometric $L_{\rm AGN} > 10^{44}$ erg s\inv\ (cyan), for mid-IR AGN selected using the \wise\ two-color cut proposed in this work (black), and for dual mid-IR AGN selected with the same color cut (red). The column density in each bin is calculated as the time-weighted median over all viewing angles, for all eight fiducial simulations; the error bars denote the median absolute deviation. The AGN column density peaks at $\ga 10^{23}$ - $10^{24}$ cm s$^{-2}$ in the late stages of the merger, at BH separations $< 3$ kpc and in the first 40 Myr post-merger. This demonstrates that AGN are preferentially obscured in advanced mergers, and that environmentally-driven obscuration can be significant during mergers even on galactic scales.
\label{fig:column_dens}}
\end{figure*}

In order to obtain an upper limit on the amount of galactic-scale obscuration, we calculate \nh\ for the {\em total} gas density along a given line of sight -- i.e., we include the cold-phase gas along with the hot phase, rather than assuming as in the radiative transfer calculations that cold-phase gas has a negligible volume filling factor. Note that, because more of the ISM is in the hot phase during the early merger stage and long after the merger (particularly in the nuclear regions), we expect the actual evolution in \nh\ throughout the merger to be even stronger than Figure \ref{fig:column_dens} suggests. 

Figure \ref{fig:column_dens} shows how the AGN  column density varies with merger stage. We consider AGN selected via bolometric luminosity ($L_{\rm AGN}>10^{44}$ erg s\inv) and via the two-color \wise\ selection criterion proposed in this work. Given our findings in \S~\ref{ssec:dual_agn}, we also examine \nh\ for \wise-selected {\em dual} AGN. In all cases, the column density along the line of sight to each BH increases in the final stage of coalescence, for BH separations $< 3$ kpc. High obscuration persists (\nh\ $\ga 10^{23}$ cm$^{-2}$) for the first $40$ Myr after the BH merger before declining significantly.  The effect is most dramatic for the dual mid-IR AGN, particularly at the smallest BH separations. (Note that in the gas-rich, major merger simulations, nearly all AGN at small nuclear separations are dual mid-IR AGN; Figures \ref{fig:wise_dual_projsep_w05} \& \ref{fig:column_dens_res}). This further emphasizes the need for high resolution imaging and spectroscopy with \jwst\ to reveal kpc-scale dual nuclei in these preferentially obscured systems.  There is substantial variation in \nh\ between mergers and for different viewing angles, but in all cases, the highest \nh\ always occurs in the bin just before or just after the BH merger. In the major, gas-rich merger simulations, we find peak \nh\ approaching the Compton-thick regime ($\ga 10^{24}$ cm$^{-2}$), which is remarkable given the limitations of finite resolution in our simulations (see also \ref{fig:column_dens_res}). 

If we consider other mid-IR selection criteria, we may expect more stringent color cuts to preferentially select AGN with higher \nh, owing to the associations between obscuration, luminosity, and redder mid-IR colors. We find, however, that during peak activity (during final coalescence and just after the merger) the AGN column densities are insensitive to the mid-IR criterion used during final coalescence and just after the merger. This reflects the strong tendency for advanced mergers to contain luminous, obscured (dual) AGN around the time of coalescence. At slightly earlier times (nuclear separations of 3-10 kpc), a more stringent \wonetwo\ $>0.8$ cut does indeed select single AGN with higher median \nh\ as expected (by a factor of a few), though the dual AGN column densities are still insensitive to the color cut. At slightly later post-merger times (40-80 Myr), however the median \nh\ is higher by a factor of ~30 for AGN with \wonetwo\ $>0.8$. 

In five of the major merger simulations, the peak column densities reach the Compton-thick regime ($> 10^{24}$ cm$^{-2}$) during coalescence, for our upper-limit calculation of $N_{\rm H}$. This is notable, given that the finite resolution of the simulations does not resolve the high density AGN ``torus" region that is expected to dominate the obscuring column in many, if not most, Compton-thick AGN. The degree of correlation between galactic-scale and torus-scale obscuration remains an important open question. And it bears repeating that our calculation is an upper limit to the the amount of galactic-scale obscuration, since we consider the total gas column density. Nonetheless, our results demonstrate that galactic-scale, environmentally-driven obscuration can be a significant contributor to AGN obscuration during major mergers. These findings agree well with recent X-ray and near-IR studies of merger-triggered AGN \citep{koss18, ricci17, satyap17,lansbu17}, in which a strong association between mergers and highly obscured and Compton-thick AGN is found in the late merger stages ($<$ 1 - 10 kpc).

\section{Summary \& Conclusions}
\label{sec:conclusions}

Using hydrodynamic simulations coupled with dust radiative transfer in post-processing, we have studied the evolution of nuclear obscuration and mid-IR AGN signatures during galaxy mergers. We focus in particular on determining the completeness and reliability of mid-IR color selection for identifying obscured (dual) AGN in advanced mergers. IR selection has revealed large populations of AGN that are hidden at other wavelengths, including a disproportionate number of AGN in ongoing mergers. Understanding the efficacy of mid-IR color selection in a variety of merger environments and merger stages is therefore crucial for leveraging large surveys, such as the {\em WISE} All-Sky Survey, to understand the long-debated role of mergers in fueling AGN.

Mid-IR selected AGN are much less sensitive to attenuation by gas and dust than AGN selected in optical or soft X-ray bands. Nonetheless, we find that much of the AGN lifetime is still missed with common mid-IR selection criteria, {\em even in the late stages of gas-rich major mergers.} The AGN must contribute at least 30 - 50\% of the total bolometric luminosity to be detected via a \wonetwo\ $> 0.8$ single-color cut, such that 40-50\% of moderate-luminosity AGN ($L_{\rm bol} > 10^{44}$ erg s\inv) would not be identified. The J11 two-color cut (using \wtwothree\ in addition to \wonetwo) has similar completeness for these AGN. 

A less stringent single-color cut of \wonetwo\ $> 0.5$ selects merger-triggered AGN with much higher completeness (median 100\% for $L_{\rm bol} > 10^{44}$ erg s\inv, with all advanced mergers $>$75\% complete). Moreover, the greater completeness of this relaxed \wise\ color cut has a minimal trade-off in the reliability of AGN selection for $z<1$. Most of our merger simulations have no contamination at all of the \wonetwo\ $> 0.5$ color cut from star formation, and only in extreme starbursts (sSFR $\ga 10^{-8.5}$ yr\inv) does star formation alone cause any significant reddening of the \wise\ \wonetwo\ color. More importantly, because nuclear starbursts and AGN are often co-spatial and nearly simultaneous during the final coalescence of merging galaxies, we find that {\em even when intense merger-triggered starbursts mimic AGN-like mid-IR colors, they are always accompanied by an AGN. }

We define a new two-color \wise\ selection criterion based on our study of AGN in merging systems (Equation \ref{eqn:mycut}). This combines the high completeness of the \wonetwo\ $> 0.5$ single-color cut with a \wtwothree\ cut that excludes pure starbursts but includes highly-obscured luminous AGN in ULIRGs. We find that this more lenient selection criterion is well-suited for mid-IR studies of AGN, particularly merger-triggered AGN, with high completeness out to $z\sim0.5$ and negligible contamination from stellar emission at $z \la 1$. 

Finally, we show that mid-IR color selection of merger-triggered AGN is also remarkably effective at identifying systems that contain {\em dual} AGN. This results from the strong correlation between AGN luminosity and mid-IR colors, and between AGN luminosity and merger stage. In other words, the most luminous AGN produce the reddest mid-IR colors, and they are preferentially found in advanced mergers that have $\la$ kpc-scale BH pairs. A \wonetwo\ $> 0.5$ cut, and the two-color cut defined in this work, select virtually {\em all} dual AGN (with $L_{1,2} > 10^{44}$ erg s\inv) throughout the merger. A majority of these dual AGN would also be identified with other commonly used selection criteria (\wonetwo\ $> 0.8$ or a two-color wedge, cf. J11). Moreover, dual AGN have the highest duty cycle at the smallest separations ($0 < a_{\rm sep} < 3$ kpc). This indicates that many dual AGN in \wise-selected samples are still unresolved. Indeed, initial X-ray and near-IR spectroscopic follow-up studies have already yielded promising results 
\citep{satyap17,elliso17}. A complementary adaptive optics follow-up study of hard X-ray selected AGN also revealed a high fraction of small-scale dual nuclei in late-stage mergers \citep{koss18}. 

Not all of the mergers in our simulation suite go through a luminous, obscured AGN phase, but this is a ubiquitous outcome in the gas-rich, major mergers. The AGN luminosity and line-of-sight column density to the BHs are well correlated, particularly in the final stages of the galaxies' coalescence when the BH fueling and obscuration both typically reach their peak. Moreover, because we do not resolve obscuration on ``torus" scales near the AGN, the peak column densities ($N_{\rm H} \ga 10^{24}$ cm$^{-2}$) suggest that galactic-scale obscuration during mergers can contribute significantly to the total attenuation of AGN emission, even in the Compton-thick regime. This stands in contrast to the standard AGN unification picture, in which viewing angle is the dominant factor in the degree of AGN obscuration.  These obscured phases coincide with a critical stage of BH growth during which the BH can more than double its mass. This supports the expectation that the most rapid BH fueling episodes should be preferentially obscured during late-stage mergers, indicating that merger-triggered AGN fueling can be easily missed. Obscured AGN in late-stage mergers are therefore ideal targets for further study in the infrared, and in particular, imaging and resolved spectroscopy with \jwst\ will soon be able to reveal merger dynamics on sub-kpc scales in unprecedented detail.

\section*{Acknowledgements}
We thank the anonymous referee for their helpful comments that improved the quality of our paper. This work used Extreme Science and Engineering Discovery Environment (XSEDE) resources at the San Diego Supercomputing Center and the Texas Advanced Computing Center through allocation AST130041. XSEDE is supported by National Science Foundation grant number ACI-1548562. L.B. acknowledges support from a National Science Foundation grant (AST-1715413). G.F.S. appreciates support from a Giacconi Fellowship at the Space Telescope Science Institute, which is operated by the Association of Universities for Research in Astronomy, Inc., under NASA contract NAS 5-26555.

\bibliography{refs_wise_agn_rev1}

\appendix

\section{Numerical Resolution}
\label{ssec:resolution}

For a subset of four gas-rich mergers (denoted in boldface in Table \ref{table:merger_models}), we run simulations with 10$\times$ higher mass resolution and $10^{1/3} \times$ higher spatial resolution. Overall, we find that the qualitative evolution of key quantities is unchanged at higher resolution, and the total amount of star formation during each simulation is consistent to within 10\%. The peak SFR, peak AGN luminosity, peak $N_{\rm H}$, and total BH mass accreted are consistent within a factor of $\sim 3$ - 4; the largest difference in peak $L_{\rm AGN}$ occurs in the D1D0 simulation, where the peak AGN luminosity is six times lower in the high-resolution version. These modest variations are to be expected given the stochastic nature of BH accretion and the fluctuations in instantaneous quantities in a highly dynamic environment. For the most part, they do not show any systematic trends. 

We do find slightly higher AGN column densities before the merger in the high-resolution runs, and significantly higher AGN column densities after the merger. This is largely because, in each case, we have calculated $N_{\rm H}$ along an aperture that corresponds to the effective spatial resolution of the simulation: 136 pc and 64 pc for the fiducial and high-resolution simulations, respectively. In other words, a smaller (and thus denser) nuclear region is being resolved in the high-resolution  simulations. This indicates that AGN may be even more preferentially obscured in mergers than our fiducial results suggest. However, when we perform similar analysis for the BH column densities at all timesteps (not just the AGN), we find that the differences in $N_{\rm H}$ at higher resolution are subdominant to the variation with galaxy merger parameters. Moreover, we find that our conclusions in this work do not depend on these variations in $N_{\rm H}$ with simulation resolution.

\begin{figure}
\begin{center}
\includegraphics[width=0.45\textwidth]{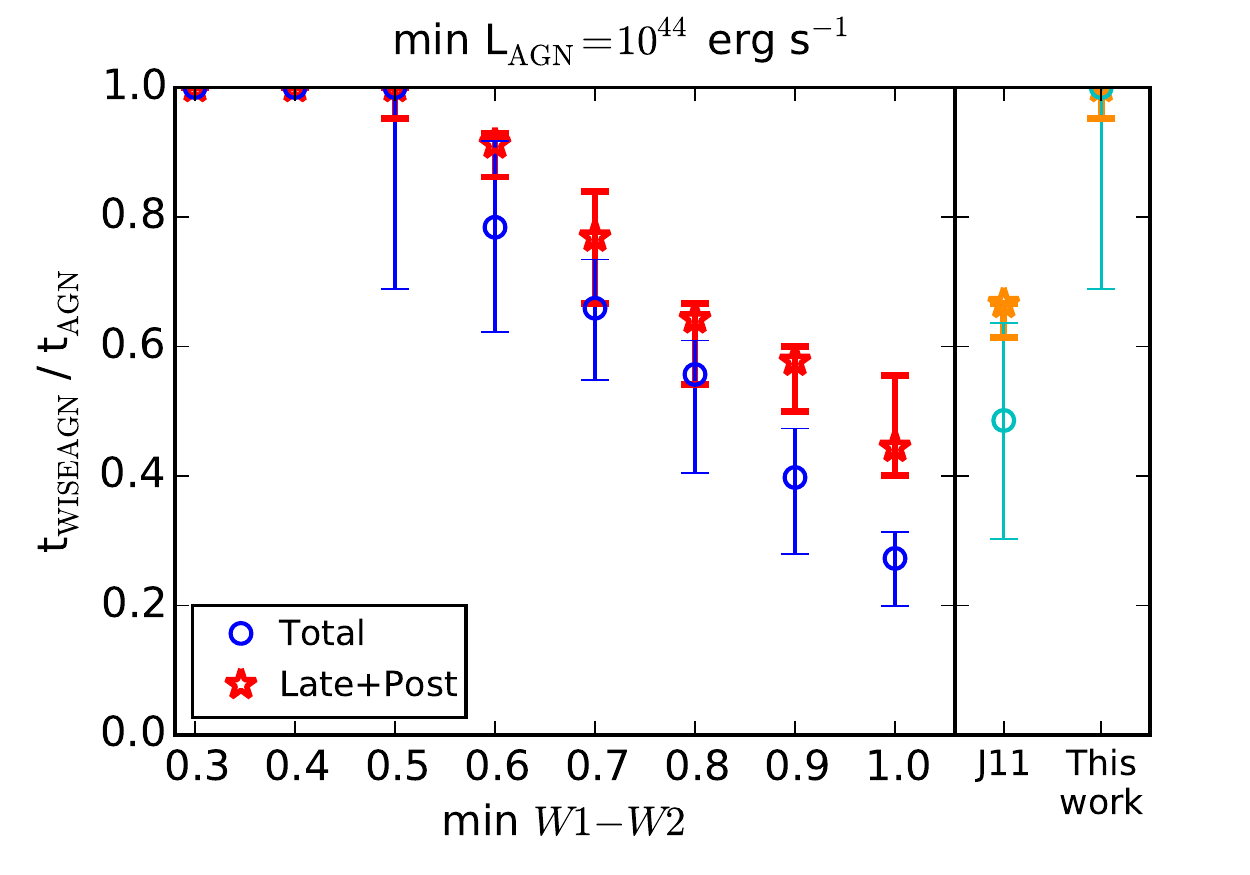}
\includegraphics[width=0.45\textwidth]{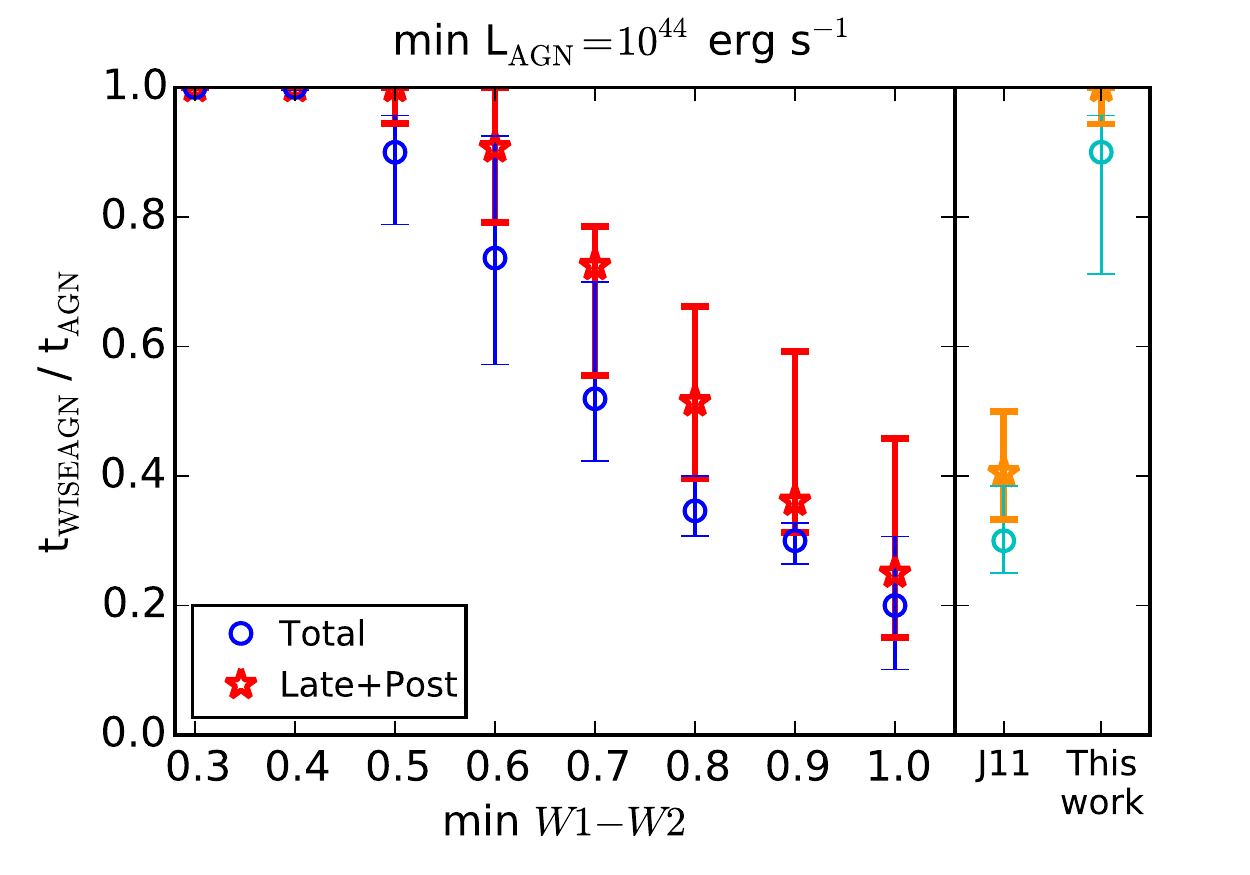}
\end{center}
\caption{As in Figure \ref{fig:wiseagn_frac_allsim}, the completeness of various mid-IR AGN selection criteria is shown, for the high-resolution simulations ({\em bottom panel}) and their fiducial-resolution counterparts ({\em top panel}). Completeness is slightly lower on average in the higher resolution simulations, owing to a small increase in dust-self absorption around buried AGN. But the mid-IR selection completeness is consistent between the high- and fiducial-resolution simulations, within the level of variation between individual simulations. 
\label{fig:wiseagn_frac_allsim_res}}
\end{figure}

Figure \ref{fig:wiseagn_frac_allsim_res} shows the completeness of mid-IR selection criteria as in Figure \ref{fig:wiseagn_frac_allsim}, for high resolution versus fiducial resolution. For the more stringent \wonetwo\ color cuts, the high-resolution simulations have marginally lower median completeness, primarily because the higher $N_{\rm H}$ increases the self-absorption of emission from the hottest, AGN-heated dust. The slight increase in dust self-absorption around the buried AGN at higher resolution produces excess emission at longer mid-IR wavelengths, and thus redder \wtwothree\ colors, which slightly lowers completeness for the two-color selection criteria as well. This is a very small effect, however; here again the differences between high and fiducial resolution are within the level of variation between simulations. 

Even in the absence of a buried AGN, the high-resolution AGNx0 simulations produce slightly more compact nuclear starbursts, which result in longer lifetimes with \wonetwo\ $ > 0.5$ from star formation alone. The fiducial and high-resolution results are still consistent within the level of variation between simulations, and the median ``contamination" for \wonetwo $>0.5$ is $\sim 0.1$ in either case. (Note that the simulation subset used in our resolution study includes two of the three fiducial mergers that exceed \wonetwo\ $ = 0.5$, so this sample has higher median contamination than the full fiducial suite. Most importantly, the ``true" contamination -- i.e., the amount of mid-IR contamination by star formation in the AGNx0 simulations when an AGN is {\em not} simultaneously present in the fiducial runs -- is still zero for all selection criteria considered in this work.

\begin{figure}
\begin{center}
\includegraphics[width=0.47\textwidth,  trim={0.15cm 0cm 0cm 0cm},clip]{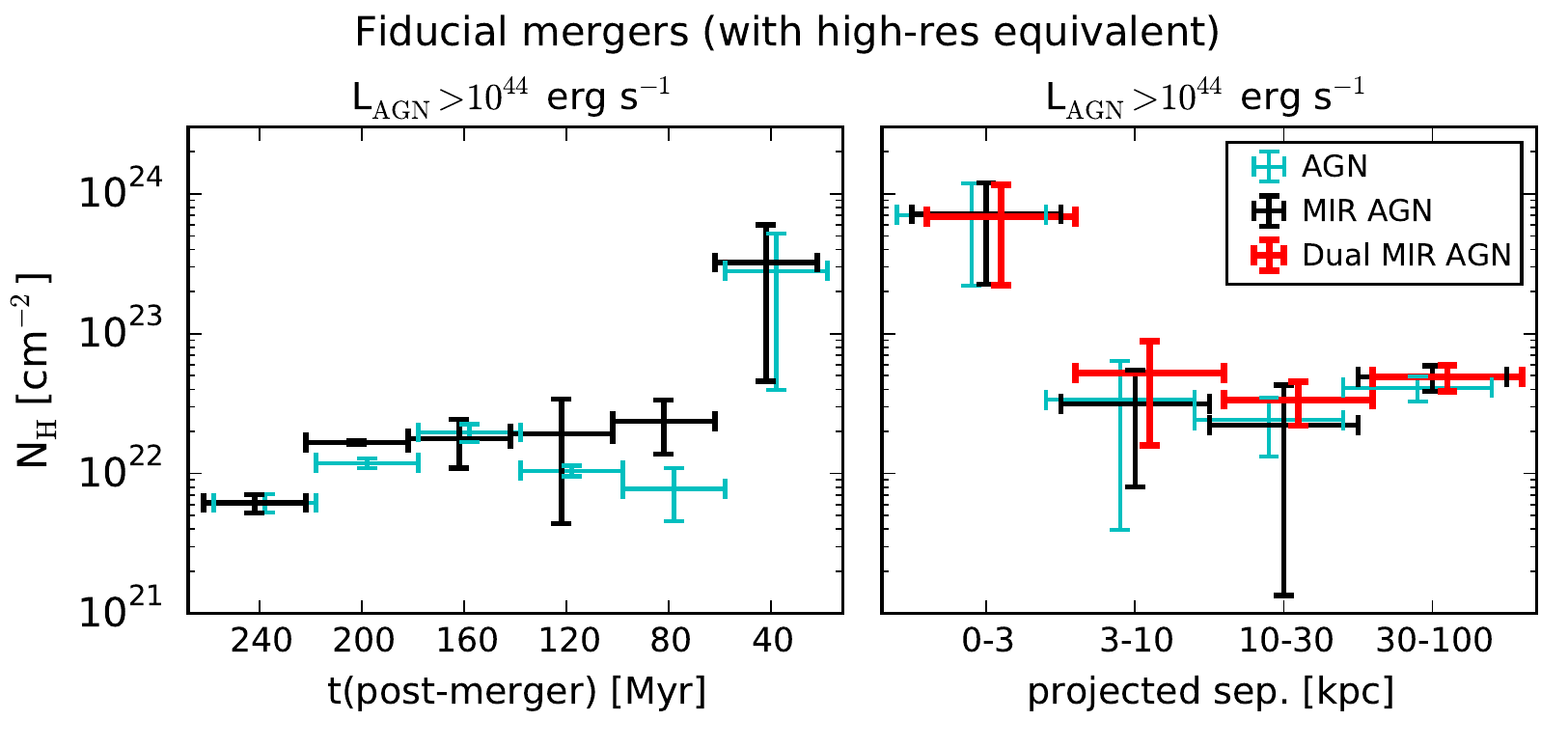}
\includegraphics[width=0.47\textwidth,  trim={0.15cm 0cm 0cm -0.5cm},clip]{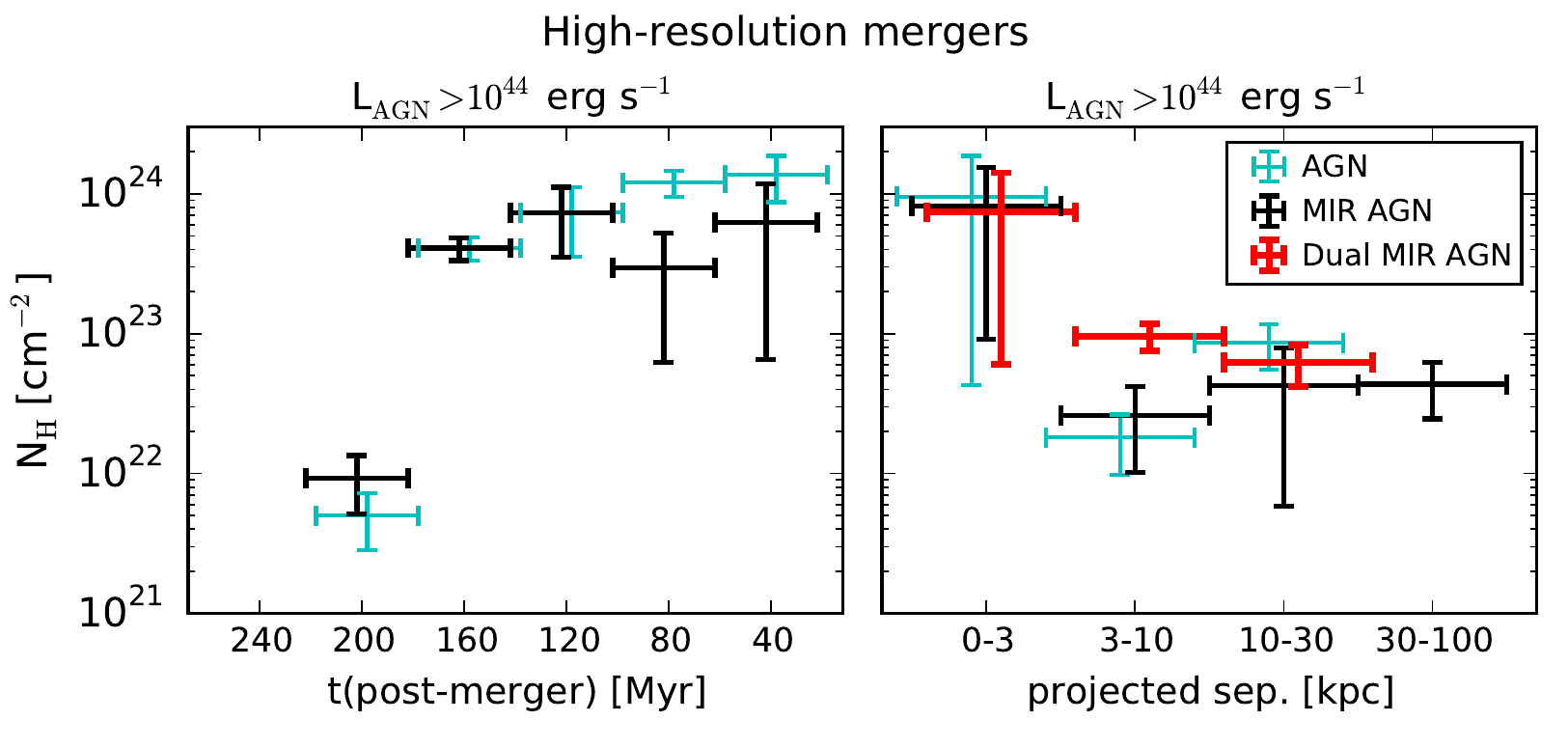}
\end{center}
\caption{Similar to Figure \ref{fig:column_dens}, the top and bottom plots show the evolution of nuclear column density $N_{\rm H}$ (along the line of sight to the AGN for the four fiducial-resolution counterparts to the high-resolution runs, and for the high-resolution simulations, respectively. As before, time evolution proceeds from right to left in these plots. A 2.2$\times$ smaller aperture is used for the high-resolution $N_{\rm H}$ calculation, to reflect the higher spatial resolution in both the {GADGET} and {SUNRISE} simulations. Although the AGN in the high-resolution runs have higher $N_{\rm H}$ after the BH merger for this subsample of gas-rich major mergers, overall the nuclear column density is consistent between the fiducial and high-resolution simulations, within the variation between individual simulations, and the differences in $N_{\rm H}$ do not affect our conclusions.
\label{fig:column_dens_res}}
\end{figure}

\end{document}